\documentclass[longauth]{aa}  

\usepackage{graphicx}
\usepackage{relsize}
\usepackage{txfonts}
\usepackage{xcolor}

\definecolor{bblue}{HTML}{002099} 
\usepackage[breaklinks,colorlinks,citecolor=bblue,linkcolor=bblue,urlcolor=bblue]{hyperref}
\usepackage[capitalise]{cleveref}
\DeclareRobustCommand{\abbrevcrefs}{\crefname{section}{Sect.}{Sects.}}
\DeclareRobustCommand{\cshref}[1]{{\abbrevcrefs\cref{#1}}} 

\usepackage[all]{hypcap} 

\begin{document} 

   \title{The miniJPAS survey:\\ AGN \& host galaxy co-evolution of X-ray selected sources}

   \author{I.E.\,L\'opez\inst{\ref{ins:difa},\ref{ins:oas}}\thanks{e-mail: ivanezequiel.lopez2@unibo.it}
        \and
        M.\,Brusa\inst{\ref{ins:difa},\ref{ins:oas}}
        \and
        S.\,Bonoli\inst{\ref{ins:dipc},\ref{ins:ikerbasque}}
        \and
        F.\,Shankar\inst{\ref{ins:southampton}}
        \and 
        N.\,Acharya\inst{\ref{ins:dipc}}
        \and
        B.\,Laloux\inst{\ref{ins:athens},\ref{durham}}
        \and
        K.\,Dolag\inst{\ref{ins:unimunich},\ref{ins:maxplanck}}
        \and
        A.\,Georgakakis\inst{\ref{ins:athens}}
        \and 
        A.\,Lapi\inst{\ref{ins:sissa}}
        \and
        C.\,Ramos\,Almeida\inst{\ref{ins:iac},\ref{ins:unilalaguna}}
        \and        
        M.\,Salvato\inst{\ref{ins:maxplanck},\ref{ins:exzellenzcluster}}
        \and
        J.\,Chaves-Montero\inst{\ref{ins:dipc},\ref{ins:barcelona}}
        \and
        P.\,Coelho\inst{\ref{ins:astrosaopaulo},\ref{ins:teruel}}
        \and
        L.A.\,D\'iaz-Garc\'ia\inst{\ref{ins:iaa}}
        \and
        J.A.\,Fernández-Ontiveros\inst{\ref{ins:teruel}}
        \and
        A.\,Hernán-Caballero\inst{\ref{ins:teruel}}
        \and
        R.M.\,González Delgado\inst{\ref{ins:iaa}}
        \and
        I.\,Marquez\inst{\ref{ins:iaa}}
        \and
        M.\,Povi\'c\inst{\ref{ins:EORC-ethiopia},\ref{ins:iaa},\ref{ins:mbarara}}
        \and
        R.\,Soria\inst{\ref{ins:beijing},\ref{ins:Sydney},\ref{ins:torino}}
        \and
        C.\,Queiroz\inst{\ref{ins:fisicasaopaulo},\ref{ins:portoalegre}}
        \and
        P.T.\,Rahna\inst{\ref{ins:CASKey}}
        \and
        R.\,Abramo\inst{\ref{ins:fisicasaopaulo}}
        \and
        J.\,Alcaniz\inst{\ref{ins:ON}}
        \and
        N.\,Benitez\inst{\ref{ins:iaa}}
        \and
        S.\,Carneiro\inst{\ref{ins:fisicabahia}}
        \and
        J.\,Cenarro\inst{\ref{ins:teruel}}
        \and
        D.\,Cristóbal-Hornillos\inst{\ref{ins:teruel}}
        \and
        R.\,Dupke\inst{\ref{ins:ON}}
        \and
        A.\,Ederoclite\inst{\ref{ins:teruel}}
        \and
        C.\,López-Sanjuan\inst{\ref{ins:teruel}}
        \and
        A.\,Marín-Franch\inst{\ref{ins:teruel}}
        \and
        C.\,Mendes~de~Oliveira\inst{\ref{ins:astrosaopaulo}}
        \and
        M.\,Moles\inst{\ref{ins:teruel}}
        \and
        L.\,Sodré\,Jr.\inst{\ref{ins:astrosaopaulo}}
        \and
        K.\,Taylor\inst{\ref{ins:instruments4}}
        \and
        J.\,Varela\inst{\ref{ins:teruel}}
        \and
        H.V.\,Ramió\inst{\ref{ins:teruel}}
}

   \institute{
            Dipartimento di Fisica e Astronomia "Augusto Righi", Università di Bologna, via Gobetti 93/2, 40129 Bologna, Italy \label{ins:difa}
            \and
            INAF - Osservatorio di Astrofisica e Scienza dello Spazio di Bologna, via Gobetti 93/3, 40129 Bologna, Italy \label{ins:oas}
            \and
            Donostia International Physics Center (DIPC), Manuel Lardizabal Ibilbidea, 4, Donostia-San Sebastián, Spain\label{ins:dipc}
            \and
            IKERBASQUE, Basque Foundation for Science, E-48013, Bilbao, Spain\label{ins:ikerbasque}
            \and
            School of Physics and Astronomy, University of Southampton, Highfield, Southampton SO17 1BJ, UK\label{ins:southampton}
            \and
            Institute for Astronomy \& Astrophysics, National Observatory of Athens, V. Paulou \& I. Metaxa 11532, Greece\label{ins:athens}
            \and
            Centre for Extragalactic Astronomy, Department of Physics, Durham University, Durham, DH1 3LE, UK\label{durham}
            \and
            Universitäts-Sternwarte, Fakultät für Physik, Ludwig-Maximilians-Universität München, Scheinerstr.1, 81679 München, Germany\label{ins:unimunich}
            \and
            Max-Planck-Institut für Astrophysik, Karl-Schwarzschild-Straße 1, 85741 Garching, Germany\label{ins:maxplanck}
            \and
            SISSA, Via Bonomea 265, 34136 Trieste, Italy \label{ins:sissa}
            \and
            Instituto de Astrofísica de Canarias, Calle Vía Láctea, s/n, E-38205, La Laguna, Tenerife, Spain\label{ins:iac}
            \and
            Departamento de Astrofísica, Universidad de La Laguna, E-38206 La Laguna, Tenerife, Spain\label{ins:unilalaguna}
            \and
            Exzellenzcluster ORIGINS, Boltzmannstr. 2, 85748 Garching, Germany\label{ins:exzellenzcluster}
            \and
            Institut de Física d’Altes Energies, The Barcelona Institute of Science and Technology, Campus UAB, E-08193 Bellaterra (Barcelona), Spain\label{ins:barcelona}
            \and
            Universidade de S\~ao Paulo, Instituto de Astronomia, Geofísica e Ciências Atmosféricas, Rua do Mat\~ao, 1226, 05508-090, S\~ao Paulo, SP, Brazil \label{ins:astrosaopaulo}
            \and
            Centro de Estudios de Física del Cosmos de Aragón (CEFCA), Unidad Asociada al CSIC, Plaza San Juan, 1, E-44001, Teruel, Spain \label{ins:teruel}
            \and
            Instituto de Astrof\'isica de Andaluc\'ia (IAA-CSIC), Glorieta de la Astronomía s/n, E-18008 Granada, Spain \label{ins:iaa}
            \and
            Astronomy and Astrophysics Research and Development Department, Entoto Observatory and Research Center (EORC), Space Science and Geospatial Institute (SSGI), P.O. Box 33679, Addis Ababa, Ethiopia\label{ins:EORC-ethiopia}
            \and
            Physics Department, Faculty of Science, Mbarara University of Science and Technology (MUST), P.O. Box 1410, Mbarara, Uganda\label{ins:mbarara}
            \and
            College of Astronomy and Space Sciences, University of the Chinese Academy of Sciences, Beijing 100049, China\label{ins:beijing}
            \and
            Sydney Institute for Astronomy, School of Physics A28, The University of Sydney, Sydney, NSW 2006, Australia\label{ins:Sydney}
            \and
            INAF - Osservatorio Astrofisico di Torino, Strada Osservatorio 20, 10025, Pino Torinese, Italy\label{ins:torino}
            \and
            Departamento de F\'isica Matem\'atica, Instituto de F\'{\i}sica, Universidade de S\~ao Paulo, Rua do Mat\~ao, 1371, CEP 05508-090, S\~ao Paulo, Brazil \label{ins:fisicasaopaulo}
            \and
            Departamento de Astronomia, Instituto de F\'isica, Universidade Federal do Rio Grande do Sul (UFRGS), Av. Bento Gon\c{c}alves, 9500, Porto Alegre, RS, Brazil \label{ins:portoalegre}
            \and
            CAS Key Laboratory for Research in Galaxies and Cosmology, Shanghai Astronomical Observatory, CAS, Shanghai, 200030, China \label{ins:CASKey}
            \and
            Observatório Nacional, Rua General José Cristino, 77, São Cristóvão, 20921-400, Rio de Janeiro, RJ, Brazil\label{ins:ON}
            \and
            Instituto de Física, Universidade Federal da Bahia, 40210-340, Salvador, BA, Brazil\label{ins:fisicabahia}
            \and
            Instruments4, 4121 Pembury Place, La Canada Flintridge, CA 91011, U.S.A\label{ins:instruments4}
            }


  
 \abstract{Studies indicate strong evidence of a scaling relation in the local Universe between the supermassive black hole mass ($M_\mathrm{BH}$) and the stellar mass of their host galaxies ($M_\star$). They even show similar histories across cosmic times of their differential terms: star formation rate (SFR) and black hole accretion rate (BHAR). However, a clear picture of this coevolution is far from being understood. 

We select an X-ray sample of active galactic nuclei (AGN) up to \textit{z}\,$=$\,2.5 in the miniJPAS footprint. Their X-ray to infrared spectral energy distributions (SEDs) have been modeled with the CIGALE code, constraining the emission to 68 bands, from which 54 are the narrow filters from the miniJPAS survey. For a final sample of 308 galaxies, we derive their physical properties, like their $M_\star$, SFR, star formation history, and the luminosity produced by the accretion process of the central BH ($L_\mathrm{AGN}$). For a subsample of 113 sources, we also fit their optical spectra to obtain the gas velocity dispersion from the broad emission lines and estimate the $M_\mathrm{BH}$. We calculate the BHAR in physical units depending on two radiative efficiency regimes. We find that the Eddington ratios and its popular proxy ($L_\mathrm{X}/M_\star$) have, on average, 0.6 dex of difference, and a KS-test indicates that they come from different distributions. Our sources exhibit a considerable scatter on the $M_\mathrm{BH}$--$M_\star$ scaling relation, and this can explain the difference between the Eddington ratios and its proxy. 

We also model three evolution scenarios for each source to recover the integral properties at \textit{z}\,$=$\,0. Using the SFR and BHAR, we show a notable diminution in the scattering between $M_\mathrm{BH}$--$M_\star$. For the last scenario, we consider the SFH and a simple energy budget for the AGN accretion, and we retrieve a relation similar to the calibrations known for the local Universe. 

Our study covers $\sim$\,1\,deg$^2$ in the sky and is sensitive to biases in luminosity. Nevertheless, we show that, for bright sources, the link between the differential values (SFR and BHAR) and their decoupling based on an energy limit is the key that leads to the local $M_\mathrm{BH}$--$M_\star$ scaling relation. In the future, we plan to extend this methodology to thousand degrees of the sky using JPAS with an X-ray selection from eROSITA, to obtain an unbiased distribution of BHAR and Eddington ratios.
    }
   \keywords{galaxies: evolution --
                galaxies: active --
                galaxies: nuclei --
                galaxies: photometry --
                quasars: supermassive black holes
            }
    \maketitle

\section{Introduction}
\label{sec:1}

Since the first discovery of a quasar in \citet{1963Natur.197.1040S}, it has been proposed that the central supermassive black hole (SMBH) and its host galaxy are somehow connected \citep{1969Natur.223..690L,1982MNRAS.200..115S,2000MNRAS.317..488S}. This co-evolutionary scenario is further supported by the strong correlations between the SMBH mass ($M_\mathrm{BH}$) and various properties of the host galaxy, such as the velocity dispersion of the bulge component, stellar mass ($M_\star$), and luminosity \citep[$L_\star$; see][for reviews]{2006ApJ...644L..21F,2009NewAR..53...57S,2013ARA&A..51..511K,2016ASSL..418..263G}, and also an anticorrelation between X-ray-to-optical flux ratio and host galaxy light concentration \citep{2009ApJ...706..810P,2009ApJ...702L..51P}.

The cosmic star formation rate (SFR) density, which peaked at \textit{z}\,$\sim$\,2--3 \citep{2014ARA&A..52..415M}, has been declining since then. On the other hand, the black hole (BH) accretion rate density, estimated from the quasar luminosity function, also peaks at \textit{z}\,$\sim$\,2--3 and then drops by more than an order of magnitude at \textit{z}\,$<$\,1 \citep{2006ApJ...652..864H,2009ApJ...690...20S}. The tight correlation between the SFR and the BH accretion rate across cosmic time \citep[e.g.,][]{2008MNRAS.388.1011M,2013MNRAS.428..421S,2015ApJ...810...74A,2015MNRAS.451.1892A,2018MNRAS.475.1887Y,2020A&A...642A..65C} suggests that the BH growth is closely linked to the star formation history (SFH) of its host galaxy.

The black hole accretion rate (BHAR) is a crucial parameter that describes the BH growth rate and the efficiency of the BH feedback \citep{2014ApJ...782...69L}. A way to normalize the BHAR for different BH masses is the Eddington ratio ($\lambda_\mathrm{Edd} = L_\mathrm{AGN} / L_\mathrm{Edd}$), which measures the luminosity produced by the active galaxy nuclei (AGN) relative to the Eddington limit ($L_\mathrm{Edd}$). The $\lambda_\mathrm{Edd}$ is an essential parameter in BH-galaxy co-evolution models, as it determines the BH feedback efficiency and the degree of self-regulation of the BH growth \citep[e.g.,][]{2004ApJ...600..580G,2005Natur.433..604D,2006ApJ...650...42L}. It is also essential to study the accretion rate and its correlation with properties of the host galaxy, such as the SFR, as this can provide insights into the feedback mechanisms that regulate the growth of both the black hole and the host galaxy \citep[e.g.,][]{2014ARA&A..52..589H,2015MNRAS.449..373D,2016MNRAS.458..816H,2019ApJ...872..168S,2020A&A...642A..65C,2021MNRAS.506.2619T}.

\hfill

Despite the importance of the BHAR, it is not easy to measure it directly due to the faintness of some accreting BHs. The accretion can have different `modes' where the efficiency to produce the observed radiation changes \citep[e.g., see][for a review]{2014ARA&A..52..589H}. Photons from the accretion can also be absorbed by gas and dust that obscure observational indicators \citep[for a recent review, see][]{2018ARA&A..56..625H}. It is also difficult to directly measure $\lambda_\mathrm{Edd}$ because of its dependence on $L_\mathrm{AGN}$ and $M_\mathrm{BH}$. Since hard X-ray photons are less affected by the obscuration, a popular approach is to use $L_\mathrm{X}$ as a proxy for $L_\mathrm{AGN}$; also $M_\star$ can be a proxy for $M_\mathrm{BH}$, and hence, $L_\mathrm{Edd}$ \citep[see][for examples of these proxies]{2009A&A...507.1277B,2017MNRAS.471.1976G}. Combined, these proxies are easier to measure than $\lambda_\mathrm{Edd}$, but they can also be subject to various uncertainties and selection biases \citep{2010ApJ...720..368X,2015ApJ...813...82R}.

An alternative method to estimate the $L_\mathrm{AGN}$ is through spectral energy distribution (SED) fitting, which can disentangle the emission from the AGN and the stellar and nebular continuum of the galaxy hosting the AGN. This method has been used to estimate the Eddington rates of AGN at various redshifts \citep[e.g.,][]{2012MNRAS.427.3103B,2014MNRAS.437.3550M,2015MNRAS.447.2085S}, and, because its strength to unravel  different types of emission, is also used to study the relation between host galaxy SFR and the AGN \citep[e.g.,][]{2018A&A...618A..31M,10.1093/mnras/stac2800}. The SED-fitting improves when multiwavelength data are available, since the emission from the AGN can be observed at different bands. Moreover, the use of narrow-band filters is particularly well suited for AGN studies, as it better constrains the stellar population of the host galaxy and therefore, the AGN component.

On the other hand, it is unclear whether the $M_\mathrm{BH}$--$M_\star$ relation is the same across the cosmic time. While the relation is well-known for the local Universe for active and inactive supermassive BHs \citep[see][for examples of recent studies]{2019MNRAS.485.1278S,2020NatAs...4..282S,2021ApJ...921...36B}, it is not clear if it holds further in time. Some authors show evidence of evolution in the relation \citep[e.g.,][]{2010ApJ...708..137M,2010MNRAS.402.2453D}. In \citet{2015ApJ...805...96S}, the authors do not find a significant change in the relation until \textit{z}\,$\sim$\,1, but they find hints of a flattener relation at higher redshifts. Studies such as \citet{2021ApJ...922..142L} and  \citet{2020ApJ...889...32S} show no significant evidence of an evolution in the relationship until \textit{z}\,$\sim$\,0.8 and 2.5, respectively. The lack of certainty also remains in large-scale cosmological simulations, where there is no agreement in the expected scaling relation at \textit{z}\,$>$\,4 \citep{2022MNRAS.511.3751H}. \citet{2011ApJ...734...92J} suggest that the relationship does not imply a physically coupled growth and \citet{2022arXiv220914526G} point that mergers shape the high end of this relationship. Nevertheless, biases cannot be ignored in these studies; for example, finding overmassive galaxies for a given BH mass at different redshifts can be dominated by observational biases \citep{2014ApJ...780..162M,2020ApJ...888...37D}, and flux-limited samples are generally biased towards higher values of $M_\mathrm{BH}$--$M_\star$ \citep[e.g.,][]{2007ApJ...670..249L,2011A&A...535A..87S}. While the debate continues, it is clear that using $M_\star$ as a proxy to estimate $M_\mathrm{BH}$ needs to be taken with caution. 

In this work, we combined a sample of X-ray-selected AGN with the narrow-band data from the miniJPAS survey. miniJPAS is an optical survey \citep{2014arXiv1403.5237B,2020arXiv200701910B} with a extensive narrow-band filters system ((for more detailed description see \cshref{subsec:21}). This survey has demonstrated adequate capacity for galaxy evolution studies through the J-spectra retrieved from the narrow band filters  \citep[e.g.,][]{2021A&A...649A..79G,2022arXiv220710101R}. AGN studies are also suitable on miniJPAS; \citet{2022arXiv220200103Q} provides a selection of quasar candidates obtained with machine learning methods, and \citet{2022arXiv220700196R} detects a double-core Ly$\alpha$ morphology on two quasars using the narrow-band images. Since our aim is to study the host galaxy and AGN properties, and particularly their inferred accretion rate distributions, we chose an X-ray selection because it is one of the least biased methods to select AGN. We obtain the $M_\mathrm{BH}$ from single-epoch spectral fitting for a subsample of sources and reliable estimates of AGN accretion rate luminosities from a detailed SED fitting. We compare the measured Eddington ratios with the proxies discussed above for the subsample of sources with both BH mass and AGN luminosities, and we discuss the main differences. We also study different possible evolutionary scenarios for the sources and compare them with local scaling relations.

The paper is organized as follows. \cref{sec:2} presents the sample and all the data used for the study. The data analysis is presented in \cshref{sec:3,sec:4}, focusing respectively on the SED method used to derive AGN and host galaxy properties and the optical spectral fitting procedure used to derive the BH masses for our targets. \cref{sec:5} describes the best fit physical properties of the AGN and host galaxies, particularly the BH accretion rate. In \cshref{sec:6}, we model the evolution of the $M_\mathrm{BH}$--$M_\star$ relation from the observed \textit{z} out to \textit{z}\,$=$\,0, and finally, in \cshref{sec:7}, we summarize our conclusions. 

As cosmological parameters we adopt H$_0$ = 67.7 km s$^{-1}$ Mpc$^{-1}$ and $\Omega_{\rm m}$ = 0.307, derived by \citet{Planck15}. The AB system will be used when quoting magnitudes unless otherwise stated. Solar masses and SFRs are scaled according to a universal \citet{2003PASP..115..763C} initial mass function.

\section{Sample selection and multiwavelength data}
\label{sec:2}
This section describes the data sets used in our analysis. In \cshref{subsec:21}, we recount the miniJPAS survey, which lies along the Extended Groth Strip (EGS) field \citep{2007ApJ...660L...1D} and provides the optical data to characterize the AGN and their host galaxies. \cref{subsec:22} shows the X-ray data available in the EGS field and the AGN selection in the X-rays. In \cshref{subsec:23}, we describe the methodology followed to obtain the intrinsic X-ray fluxes. Finally, in \cshref{subsec:24} we explain the available data in other bands and the optical spectra for our source selection. 

\begin{figure*}
    \resizebox{\hsize}{!}{\includegraphics{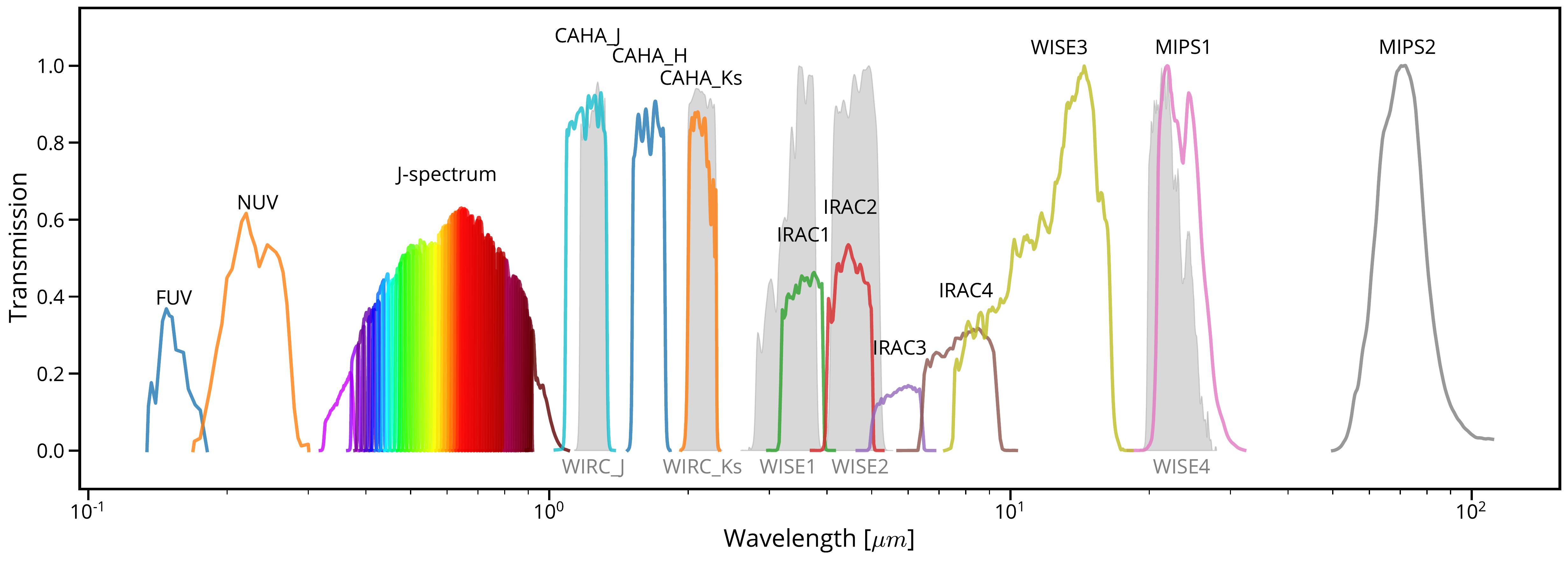}}
    \caption{In color, the main filter coverage between UV and mid-IR. For our SED fitting we have a total of 68 filters (colors). The 5 Filters shown on grayscale on the background are the second option in case the main does not have an observation on the target.}
    \label{fig:Filters}
\end{figure*}

\subsection{Narrow-band data from the miniJPAS survey}
\label{subsec:21}

miniJPAS \citep{2020arXiv200701910B} is a small proof-of-concept survey carried out by the Javalambre Physics of the Accelerating Universe Astrophysical Survey\footnote{\smaller \url{j-pas.org}} (J-PAS) collaboration \citep{2014arXiv1403.5237B}. Observations have been obtained with an interim camera mounted on the 2.55~m telescope of the Observatorio Astrofísico de Javalambre  (OAJ), and they cover a field of $\sim$\,1 deg$^2$ along the EGS field. The entire field has been observed with all the 56 optical filters of J-PAS: 54 narrow-band filters (full width at half maximum  - \textit{FWHM}\,$\simeq$\,145~\AA ) that cover the wavelength range from 3780 to 9100~\AA , and two broader filters in the blue and red wings that extend the range to 3100--10\,000~\AA . The coverage of these narrow filters is shown on \cref{fig:Filters}, in addition to the other wavelengths used in this work (see Section \cref{subsec:24} for more details).

The J-PAS filter system effectively provides a low resolution \textit{pseudo}-spectrum (from now on, J-spectrum) for every detected source and is particularly suited to study AGN \citep{2012MNRAS.423.3251A}. In the $\sim$\,1 deg$^2$ of the sky covered until now, the miniJPAS catalog contains more than 64\,000 sources detected in the \textit{r} band. This catalog is 99\% complete up to $r$ = 23.6 for point-like sources and up to \textit{r}\,$=$\,22.7 for extended sources  \citep{2020arXiv200701910B}. Point-like sources are defined as having \verb CLASS_STAR  $>0.9$ in the morphological classification from \verb SExtractor  \citep{1996A&AS..117..393B}. Considering that we used the photometric data until \textit{r}\,$<$\,23.6, in this work, we will not provide morphological information about our sources. 

miniJPAS offers different catalogs: single and dual modes. In single mode, the detection of sources is independent for each filter. This mode can be advantageous for obtaining information on faint sources with emission lines with a high signal-to-noise ratio (SNR). Because we are interested in obtaining a well-described shape of the optical SED, we used the dual-mode. In this catalog, the detection is performed in a reference band ($r$ band), and the photometry of all other filters is forced to the reference aperture (fixed shape and centroid). From the dual catalog of miniJPAS we also picked two different photometries: \verb AUTO  and \verb PSFCOR . The difference between them is that \verb AUTO  gives the magnitude within a Kron aperture, while \verb PSFCOR  is obtained in a smaller aperture and takes into consideration the differences in the point spread functions (PSFs) between the different filters \citep[for details on the photometries definition, see][]{2021A&A...654A.101H}. In \cshref{sec:3}, we discuss and explain the choice of working with \verb AUTO . We corrected all miniJPAS magnitudes for galactic extinction using the color excess $E(B$--$V)$ calculated from Bayestar17 \citep{2018MNRAS.478..651G} for each filter \citep[for details, see][]{2019A&A...631A.119L}.

\subsection{X-ray selection}
\label{subsec:22}

We selected the sources in the X-rays because it is an efficient method to search, in a wide range of redshifts, AGN with different luminosities that can be missed in other bands in the cases where the host galaxy dominates that flux  \citep[see][for a review]{2015A&ARv..23....1B}. In the past years, the EGS field has been studied with profound X-ray observations from Chandra \citep{2009ApJS..180..102L,2015ApJS..220...10N} and shallower and wider observations from XMM-Newton \citep{2020ApJS..250...32L}. All sources detected during these observations have been cataloged. The X-ray catalogs provide X-ray fluxes in the soft (0.5--2 keV) and hard (2--10 keV) bands. 

\begin{figure}[!t]
    \resizebox{\hsize}{!}{\includegraphics{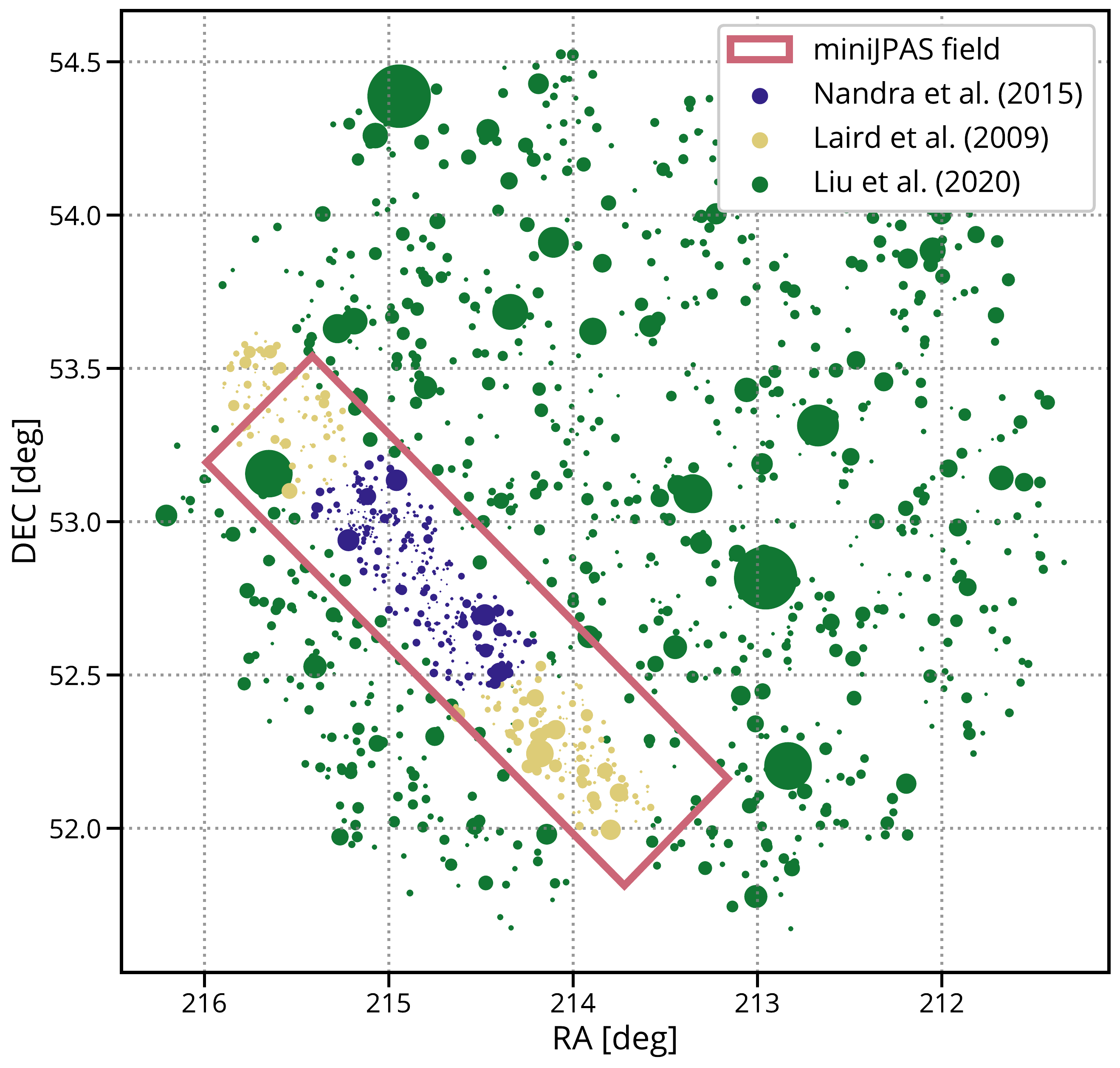}}
    \caption{Sky map of the X-ray sources on and around the miniJPAS footprint (red box). Each dot represents an X-ray source in our compiled X-ray catalog, color-coded to show its original catalog. The size of the dots is proportional to their total X-ray flux measured in 0.5--10~keV.}
    \label{fig:Sky-foot}
\end{figure}

We compiled these data in a unique catalog, keeping the deepest observations for the sources with multiple detections. For the two Chandra catalogs, we crossmatched sources within two arcsec. Since XMM-Newton has a lower spatial resolution than Chandra, we used five arcsec of maximum separation for the crossmatch between Chandra and XMM-Newton sources. We also removed the spurious sources detected in \citet{2009ApJS..180..102L} following \citet{2015ApJS..220...10N} and spectroscopically  confirmed stars. Finally, we obtain a catalog of 4928 unique X-ray sources detected in $\sim$\,6 deg$^2$ around the EGS field (1617 sources with Chandra observations and 3311 for XMM-Newton). \Cref{fig:Sky-foot} shows a sky map of the X-ray compiled catalog and the miniJPAS footprint. The original catalogs also provided reliable counterparts, obtained using likelihood estimation analysis and deep optical/IR photometric data \citep[for details, see][]{2009ApJS..180..102L,2015ApJS..220...10N,2020ApJS..250...32L} and only $\sim$\,2$\%$ of the sources do not have any optical/IR counterpart. 1661 of our 4928 unique X-ray sources lie within the miniJPAS footprint and have a reliable optical/IR counterpart.

We crossmatched these counterparts with the miniJPAS dual-mode catalog, up to \textit{r}\,$<$\,23.6 mag, obtaining 741 matches. When available, we added a confident spectroscopic redshift value from DEEP2 DR4 \citep{2013ApJS..208....5N} and SDSS DR16 \citep{2020ApJS..249....3A} using the optical/IR position and searching within a radius of one arcsec. We found robust redshift values for 430 of them (i.e. \verb ZQUALITY  $\ge$\,3 for DEEP2 and \verb zWarning  =\,0 for SDSS). We also excluded sources with any type of flag in all the miniJPAS narrow filters. These flags can be from the extraction process (close neighbor, saturated pixel, too close to a boundary, tiles overlap, between others) or because the images were affected by different technical problems in the CCD or telescope \citep[see][for details on flags]{2020arXiv200701910B}. Finally, we are left with 370 sources with X-ray fluxes, optical photometry, and a reliable redshift value. In \cref{table:nro_cuts}, we show the numbers of sources in detail for each cut. The selection done is generous to include all types of sources, but we will exclude a posteriori sources whose light is dominated by the galaxy host or the AGN, and thus the determination of their physical parameters is unreliable.

\begin{table}
\caption{Total X-ray sources in the EGS fields and our sample selection.}
\label{table:nro_cuts}
\centering
\begin{tabular}{l c c c}
\hline\hline
& All & With $z_\mathrm{spec}$ & \textit{z}\,$<$\,2.5 \\
\hline
X-ray selection &  4928 & 1394 & 1282 \\
\ \& \ inside miniJPAS area & 1661 &     532 & 507 \\
\ \& \ miniJPAS detection & 741 & 430 &    406 \\
\ \& \ miniJPAS flagged & 641 &     370 & 347 \\
\hline                                  
\end{tabular}
\end{table}

\subsection{X-ray flux correction}
\label{subsec:23}

The X-ray photons suffer a photoelectric absorption that can be modeled depending on the hydrogen column density ($N_\mathrm{H}$). This absorption can be intrinsic to the source, occurring before the photons escape the host galaxy, or local, due to the interstellar medium (ISM) in the Milky Way (MW).

Since the X-ray AGN photons originate from a non-thermal process, they can be modeled with power-law spectra, and we can predict the loss of photons for a given power-law index, redshift, and intrinsic $N_\mathrm{H}$. Because the response curve of each X-ray telescope is different and can change during its useful life, this relation also depends on the instrument and date of observation. 

\begin{figure}[!ht]
    \resizebox{\hsize}{!}{\includegraphics{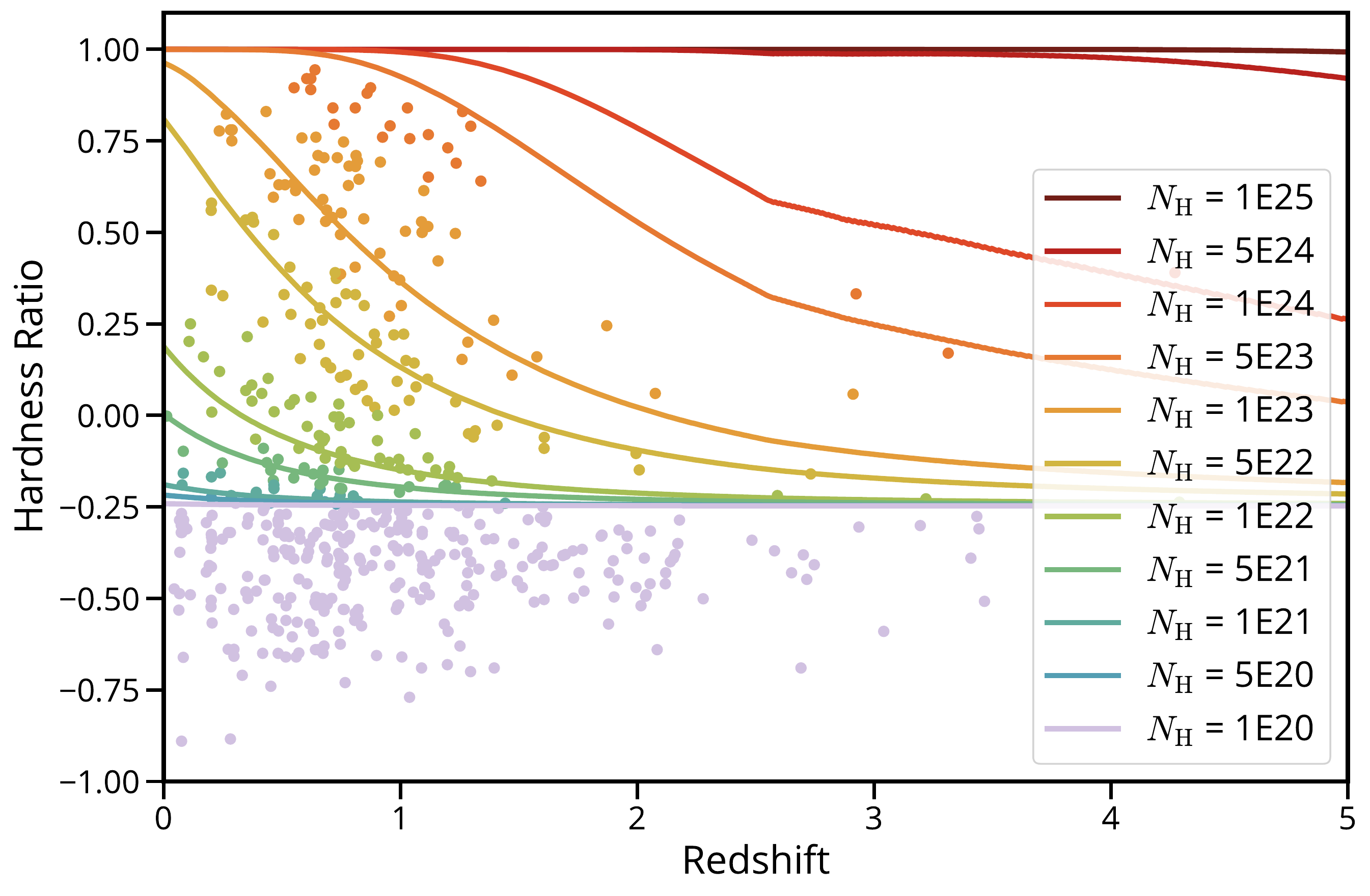}}
    \caption{Hardness ratio as a function of redshift for Chandra sources in our sample (dots). The solid lines show the value of the corresponding column density, $N_\mathrm{H}$, for a fixed $\Gamma$\,$=$\,1.4. The color of each source corresponds to the assigned $N_\mathrm{H}$ (values in cm$^{-2}$).}
    \label{fig:NH-HR}
\end{figure}

To estimate $N_\mathrm{H}$, we used the Hardness Ratio, HR $=\frac{\mathrm{H-S}}{\mathrm{H+S}}$, where H and S are counts in the soft and hard bands, respectively. H is measured in the 2--10 keV band, while S is in the 0.5--2 keV band. We use the software PIMMS\footnote{\smaller \url{https://heasarc.gsfc.nasa.gov/docs/software/tools/pimms.html}} to predict how HR changes with redshift at a fixed $N_\mathrm{H}$ and photon index\footnote{Since the original catalogs do not have information to obtain fluxes from the photon counts, we used the original photon index, $\Gamma$\,$=$\,1.4 for Chandra catalogs and $\Gamma$\,$=$\,1.7 for XMM/Newton catalog.} ($\Gamma$) for Chandra and XMM-Newton main cameras, and the representative observation date for each log. As an example, we show these predictions for Chandra sources with lines in \cref{fig:NH-HR}. A similar approach to obtain $N_\mathrm{H}$ from HR was employed in \citet{2016ApJ...817...34M}.

We computed the bayesian HR using the program Bayesian Estimation of Hardness Ratios \citep[BEHR;][]{2006ApJ...652..610P} for all the sources (shown as dots in \cref{fig:NH-HR}). Finally, we selected the nearest curve for each source, estimating the closest value of intrinsic $N_\mathrm{H}$ for them. 

\begin{figure}
    \resizebox{\hsize}{!}{\includegraphics{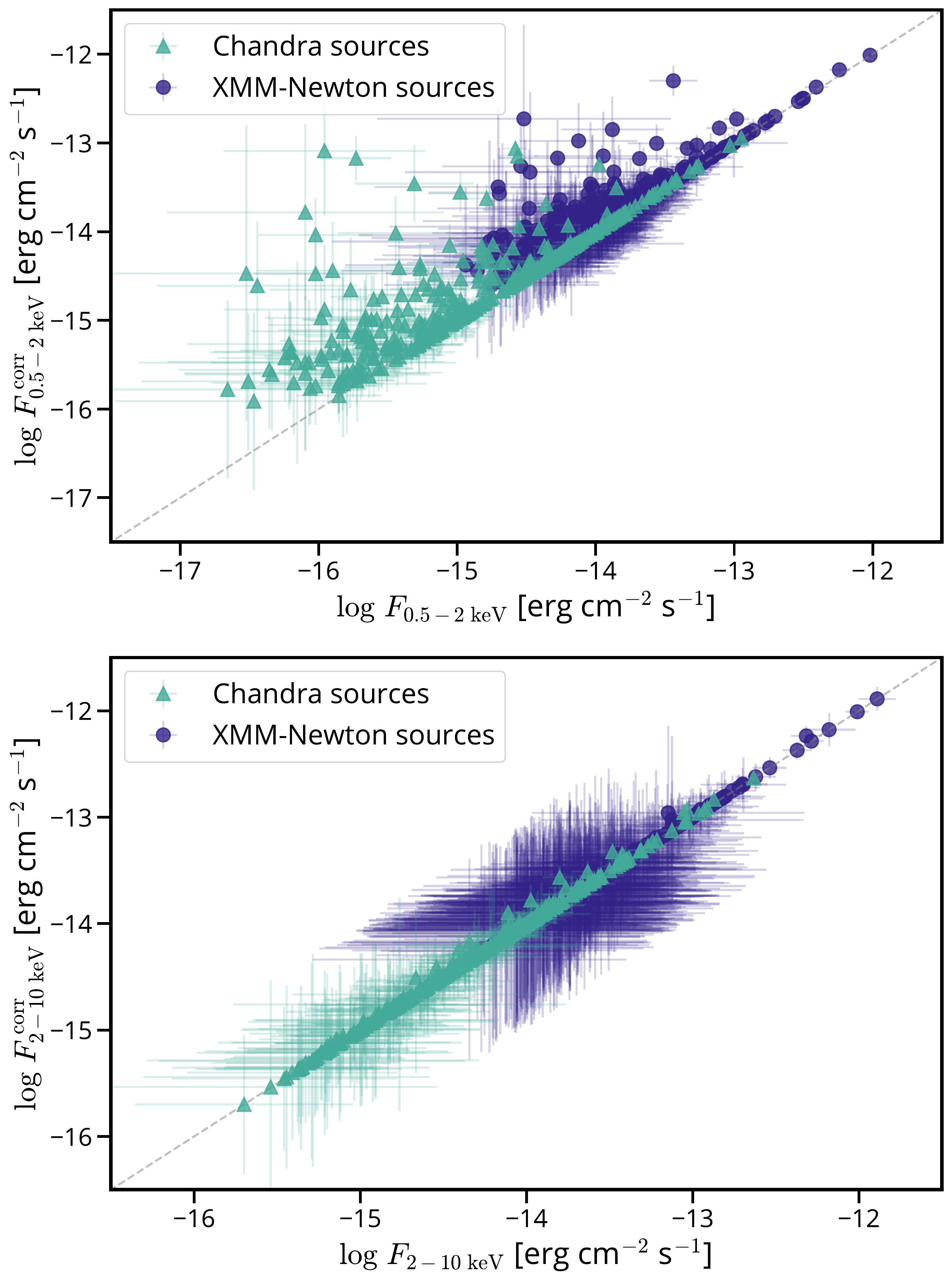}}
    \caption{Intrinsic X-ray fluxes as inferred after the $N_\mathrm{H}$ correction vs. measured fluxes in the soft (0.5-2 keV, upper panel) and hard (2-10 keV, lower panel) bands. }
    \label{fig:Xray-corr}
\end{figure}

The flux correction for the MW absorption was already performed in the original catalogs. Then we just apply the correction for the intrinsic absorption to obtain the intrinsic values of X-ray flux on the soft and the hard bands. We use PIMMS, adopting the estimated intrinsic $N_\mathrm{H}$. In Fig. \ref{fig:Xray-corr}, we show the flux corrected for intrinsic absorption following the procedure outlined above as a function of the detected flux for the soft and the hard bands. As expected, absorption affects the hard band less than the soft band.

\begin{figure}
    \resizebox{\hsize}{!}{\includegraphics{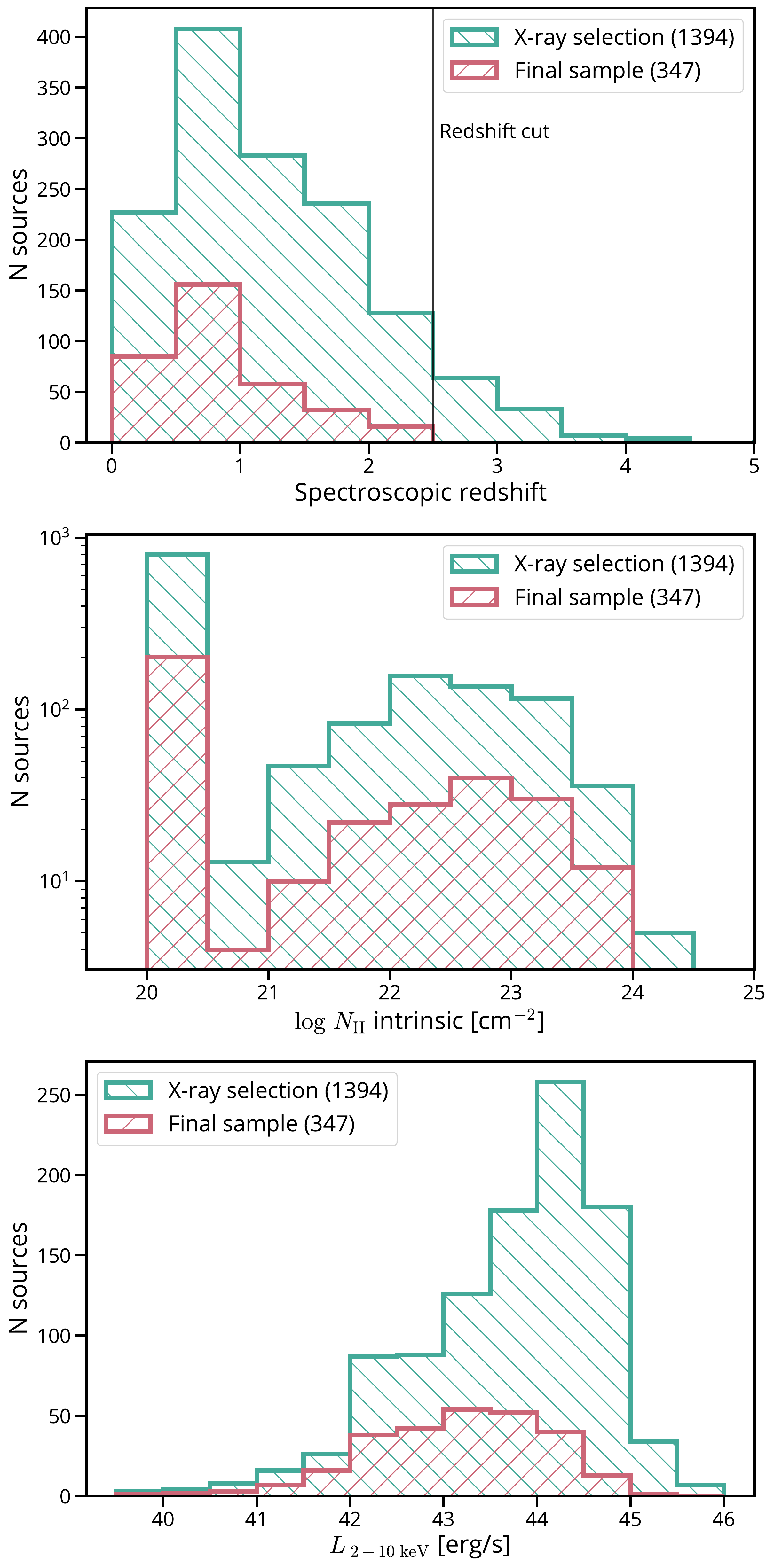}}
    \caption{Distributions of spectroscopic redshifts (upper panel), intrinsic column density $N_H$ (middle panel), and hard X-ray luminosity (lower panel) for all sources in the EGS sample (green) and our final sample with miniJPAS detection and good photometry used in this work (red).}
    \label{fig:Hist}
\end{figure}

In \cref{fig:Hist}, we show the redshift distribution for all the X-ray sources with spectroscopic redshift measurements in the EGS field (1394). This distribution drops significantly after \textit{z}\,$=$\,1.5, showing a small number of sources after \textit{z}\,$=$\,3. We decided to cut in \textit{z}\,$=$\,2.5 our sample with miniJPAS detection (370) to avoid spreading our sample at higher redshifts with few sources (see \cref{table:nro_cuts}). This cut defines our final sample: 347 X-ray sources with spectroscopic redshifts and miniJPAS photometry (flagged). This sample is shown in red in \cref{fig:Hist}. In this figure, we also show the histogram for the estimated intrinsic $N_\mathrm{H}$ and the distribution of X-ray absorption-corrected luminosities for all the sources in the EGS field and our final sample. While the distribution of $N_\mathrm{H}$ is similar for both, our sample does not resemble the shape of the $L_\mathrm{X}$ distribution for the complete EGS field. Values can be found in \cref{table:t_obsprop}.

\begin{table*}
\caption{\label{table:t_obsprop}Sources analized in this work.}
\centering
\begin{tabular}{cccccccc}
\hline\hline
id & id miniJPAS & RA & DEC & redshift & $r$-band & $\log$ L$_\mathrm{X}$ & $\log N_\mathrm{H}$\\
&&[deg]&[deg]&&[mag]&[erg s$^{-1}$]&[cm$^{-2}$]\\
\hline\\
aegis\_019 & 2241-15772 & 214.610 & 52.472 &  0.681 &     23.14 & 42.86 &         20.0 \\
aegis\_021 & 2241-19043 & 214.424 & 52.473 &  1.148 &     21.88 & 44.28 &         22.0 \\
aegis\_022 & 2241-15294 & 214.626 & 52.478 &  1.993 &     22.06 & 43.56 &         22.7 \\
aegis\_026 & 2241-14038 & 214.679 & 52.489 &  1.083 &     22.03 & 43.35 &         20.0 \\
aegis\_029 & 2241-15867 & 214.568 & 52.495 &  1.605 &     20.22 & 44.11 &         20.0 \\
aegis\_032 & 2241-17939 & 214.439 & 52.498 &  0.873 &     22.88 & 43.19 &         23.7 \\
aegis\_035 & 2241-12772 & 214.755 & 52.506 &  0.238 &     19.48 & 43.53 &         21.0 \\
aegis\_036 & 2241-19320 & 214.353 & 52.507 &  0.482 &     20.14 & 42.48 &         20.0 \\
...&...&...&...&...&...&...&...\\
\hline
\end{tabular}
\tablefoot{Full table available online in digital format.}
\end{table*}

\subsection{UV/IR data and optical spectra}
\label{subsec:24}

In order to build a complete SED and to fit diverse host galaxy and AGN models, it is necessary to have multiwavelength data. Because of this, we included in our analysis all the available photometric data from ultraviolet (UV) and infrared (IR) full-sky coverage surveys when were available (\cref{fig:Filters}). The chosen filters cover UV to mid-IR with up to 68 bands as detailed below. This selection was made to cover the rest-frame UV to near-IR fluxes up to \textit{z}\,$=$\,2.5 for all sources. This will allow a good estimate of the SED, especially on the host galaxy emission.

For UV, we crossmatched our catalog with GALEX GR6/7 \citep{2014AdSpR..53..900B} within a radius of five arcsec. This chosen radius considers the PSF and astrometry accuracy of the instrument. We added the fluxes in the near UV (1350--1750~\AA) for 257 sources and in the far UV (1750--2800~\AA) for 207 sources in our final sample. This gave us a good estimation of UV photons from the sources for a majority of our sample ($\sim$\,80\%), and only 18\% of our close sources (\textit{z}\,$<$\,0.5) do not have rest-frame UV data. We corrected the UV fluxes for galactic extinction using the coefficients from \citet{2013MNRAS.430.2188Y}.

We used the J, H, and Ks bands from \citet{2008AJ....136.1325M} to cover the near-IR range.
The ALHAMBRA near-IR survey covered different fields of interest across the sky, and in particular, the ALHAMBRA-6 field overlaps with our field. We found photometry on all these bands for 120 sources of our sample. For the sources without an ALHAMBRA detection, we searched on the Palomar WIRC original AEGIS catalogs \citep{2007ApJ...660L...1D}. We added J and Ks photometry from this catalog for 67 and 178 sources, respectively. Overall, we have at least 298 sources ($\sim$\,87\%) with some flux on the NIR bands.

For the mid-IR, we used the Spitzer IRAC and MIPS photometry from \citet{2011ApJS..193...13B}. The four filters from IRAC gave us coverage between 3 to 10 microns. We found photometry for 273 of our sources for IRAC1 and 2, while 271 for IRAC3 and 272 for IRAC4. In the case of MIPS, we included 24 and 70 $\mu$m photometry for 251 and 135 sources, respectively.. For the ones without observations made from Spitzer, we used CatWISE2020 \citep{2021ApJS..253....8M} to obtain the fluxes for the W1 and W2 bands (3.4 and 4.6 $\mu$m). To get the fluxes for W3 and W4 (12 and 22 $\mu$m), we used AllWise \citep{cutri2013explanatory}. We did a color correction following the recommendation by All-WISE website\footnote{\smaller \url{wise2.ipac.caltech.edu/docs/release/allsky/expsup/sec4_4h.html}}. We used the published color correction from \citet{2010AJ....140.1868W}, and the observed color W2-W3 to estimate the power-law index for each source and applied that color correction when the magnitudes were converted into fluxes. Upper limits were added for undetected sources. In the case of W3, we also consider this filter for 262 sources ($\sim$\,75\% of the total sample) since this filter is in the gap between IRAC and MIPS (see \cref{fig:Filters}). In total, we have at least 346 sources with some photometry between 3-10 $\mu$m, and 322 with photometry at 22-24 $\mu$m.

We also searched for available spectra for our optical counterparts of each source within one arcsec. We used the spectra in the SDSS DR16 public archive\footnote{\smaller \url{www.sdss.org/dr16/spectro/}} and from the DEEP2 survey \citep{2013ApJS..208....5N}. In the case of SDSS, we found that 101 sources have at least one spectrum. For sources with more than one SDSS spectrum, these were stacked to improve the SNR, obtaining a median spectrum for each source. For DEEP2 data, we used the 1-d spectra, obtained throughout a variant of Horne optimal extraction \citep[see][for details]{2013ApJS..208....5N}, for 111 sources. Since the DEEP2 spectra are not flux calibrated, we corrected them, considering the CCD sensitivity\footnote{\smaller \url{https://www2.keck.hawaii.edu/inst/deimos/deimos_ccd_qe.html}} as a function of wavelength. With that correction, we can better recover the correct shape of the spectra. This region of the sky was also targeted with MMT \citep{2009ApJ...701.1484C,Yan_2011}. The authors shared with us their reduced spectra for 111 sources in our sample. Both SDSS and MMT spectra were flux calibrated. Considering all these spectra, we found at least one spectrum for 269 miniJPAS sources. Details on the final spectra and their analysis can be found in \cshref{sec:4}.

\section{Data Analysis: spectral energy distributions}
\label{sec:3}
The multiwavelength emission of galaxies can provide hints about their principal components: stars, dust, and gas, among others. Modeling the SEDs with different templates allows us to measure the physical properties of the host galaxy, disentangling the different components. Since our sources are active galaxies, we must also consider their nuclear emission. To perform the SED fitting, we used Code Investigating GALaxy Emission  \citep[CIGALE\footnote{version: 2022.1 - \smaller \url{https://cigale.lam.fr/}};][]{2005MNRAS.360.1413B,2009A&A...507.1793N,2019A&A...622A.103B} with the X-ray module added by \citet{2020MNRAS.491..740Y} that makes it possible to include an AGN component in the X-rays.

CIGALE is a solid SED fitting code prevalent in galaxy evolution analyses and has become more popular in the last years in AGN studies. The new features incorporated in the last update added the ability to consider the extinction of UV-optical from polar dust and the X-ray photons \citep[for details, see][]{2020MNRAS.491..740Y}. Recent works established CIGALE's efficiency in recovering specific physical parameters of the host galaxy and AGN. \citet{2021A&A...653A..70M} used a set of mocks AGN and demonstrated that CIGALE could disentangle the AGN/host emission finding an agreement between the true values of SFR and $M_\star$ and those recovered from the fitting. They also used an X-ray-selected sample with X-ray to far-IR photometry. They showed that CIGALE is powerful enough to correctly classify between type I and type II AGN, considering inclination and polar dust. Some parameters' accuracy can be improved by adding more bands; e.g., SFR is more robust when far-IR photometry is included. In our case, we did not include Herschel data in the fitting because the available data in the field was not deep enough. Even without far-IR photometry, CIGALE can obtain reliable SFR for X-ray-selected sources with spectroscopic redshift using the rest of photometric data \citep{2018A&A...618A..31M}.

For our work, we construct the SED of each source using the redshift and photometric fluxes from all the available bands described in \cref{sec:2} and \cref{fig:Filters} (2--10 keV, 0.5--2 keV, FUV, NUV, the 56 miniJPAS optical filters, J, H, Ks, IRAC1-4, WISE3, MIPS1, MIPS2). For sources without detection in IRAC bands, we used the WISE bands (W1, W2, W4). In particular, for X-rays, CIGALE requests intrinsic fluxes. We set the upper limits for non-detected bands following the completeness studies from their original catalog. This wavelength coverage allows us to build a good rest-frame SED for the AGN and host galaxy, even at redshift 2.5.

CIGALE uses independent modules that model a unique physical feature or process. For each parameter of these modules, CIGALE builds a prior from a given grid of parameters. To choose the modules and the grid of parameters, we followed  \citet{2021A&A...653A..70M} because of the similarity of our sources. In \cshref{subsec:31,subsec:32}, we describe each module used for the host galaxy and AGN and the values adopted for the parameters. A full description of modules and parameters used as input is given in Table \ref{table:cigale-input}. CIGALE also estimates two values for each output parameter: one from the best-fit model (called \textit{best} value) and another one that weighs all grid models (called \textit{bayesian} value). These weights are based on the Bayesian likelihood $\exp{(\chi^2/2)}$ associated with each model.

\subsection{Host galaxy emission}
\label{subsec:31}
For the stellar component, we use a $\tau$-delayed SFH. This parametrization is very versatile because it allows a smooth SFR with a similar shape as the average SFR density across cosmic time \citep{2014ARA&A..52..415M} and depends on the time at which the SFR peaks ($\tau$) for each source. The functional form is SFR$(t)$\,$\propto$\,$t~\tau^{-2} \exp(-t/\tau)$, and after the maximum at $t$\,$=$\,$\tau$, the SFR smoothly declines. We also include the possibility of a recent burst following \citet{2018A&A...620A..50M}. The stellar templates are from \citet{2003MNRAS.344.1000B} and an initial mass function from \citet{2003PASP..115..763C}, with a fixed solar metallicity to avoid degenerations.
The stellar emission is attenuated following the \citet{2000ApJ...533..682C} law, and the dust emission is modeled with the template from \citet{2014ApJ...784...83D}. Since our photometric data includes narrow filters, emission lines typical of star-forming regions can be detected \citep{2022A&A...661A..99M}. Because of that, we added a model for nebular gas that uses nebular templates from \citet{2011MNRAS.415.2920I}, choosing a width of 300 km~s$^{-1}$ for narrow emission lines. CIGALE also includes the possibility for low-mass and high-mass X-ray binaries (LMXB and HMXB) to fit the X-ray emission.

\subsection{AGN emission}
\label{subsec:32}

For the active nuclei, we use the Skirtor model included in X-CIGALE \citep{2020MNRAS.491..740Y}. We followed \citet{2021A&A...653A..70M} to model the different obscuration for Type I and Type II AGN, setting two possible inclinations (30 and 70 degrees) and a grid of values for the polar dust. We also set two posibilities for the torus optical depth at 9.7 $\mu$m (3.0 and 7.0).

The AGN fraction, frac$_\mathrm{AGN}$, can be used to compare the emission of the host galaxy versus the AGN. This parameter is the fraction of the total IR emission from the AGN. We used a grid to cover the possible values (0.01, 0.1, 0.2, ..., 0.9, 0.99) and to leave the possibility of obtaining a SED entirely dominated by the host galaxy or the AGN. 


The X-ray module helps to constrain the UV emission from the accretion disk using the $\alpha_{ox}$--$L_\mathrm{2500\AA}$ relation, and for it, we set an ample grid for possible values of $\alpha_{ox}$ \citep[$-1.9$, $-1.75$, ..., $-1.15$, $-1.0$;][]{2011ApJ...739...64X,2016ApJ...819..154L}. Considering that our X-ray fluxes are corrected by intrinsic absortion, we set a photon index typical for AGN of $\Gamma$\,$=$\,1.8.

\begin{table*}
\caption{Parameters and values for the modules used with CIGALE.} 
\centering
\setlength{\tabcolsep}{1.mm}
\begin{tabular}{cc}
       \hline
Parameter &  Model/values\\
	\hline\\[-1.5ex]
\multicolumn{2}{c}{Star formation history: delayed model and recent burst} \\[0.5ex]
Age of the main population & 500, 1000, 3000, 5000, 7000 Myr \\
e-folding time & 500, 1000, 3000, 5000, 7000 Myr \\ 
Age of the burst & 20, 200 Myr \\
e-folding time of the burst & 50 Myr \\
Burst stellar mass fraction & 0.0, 0.1 \\
\hline\\[-1.5ex]
\multicolumn{2}{c}{Simple Stellar population: Bruzual \& Charlot (2003)} \\[0.5ex]
Initial Mass Function & Chabrier (2003)\\
Metallicity & 0.02 (Solar) \\
\hline\\[-1.5ex]
\multicolumn{2}{c}{Galactic dust extinction} \\[0.5ex]
Dust attenuation recipe & modified Calzetti et al. (2000)\\
E(B-V)$_{young}$ &  0.0, 0.1, 0.25, 0.5, 0.75, 0.9\\
UV$^\lambda_{bump}$ & 217.5 nm\\
\hline\\[-1.5ex]
\multicolumn{2}{c}{Galactic dust emission: Dale et al. (2014)} \\[0.5ex]
$\alpha$ slope in $dM_{dust}\propto U^{-\alpha}dU$ & 1.0, 1.5, 2.0, 2.5, 3 \\
\hline\\[-1.5ex]
\multicolumn{2}{c}{Nebular} \\[0.5ex]
$\log U$ & -2.0 \\
Width lines & 300 km/s\\
\hline\\[-1.5ex]
\multicolumn{2}{c}{AGN module: SKIRTOR} \\[0.5ex]
Torus optical depth at 9.7 microns $\tau _{9.7}$ & 3, 7 \\
Torus density radial parameter $p$ & 1.0 \\ 
Torus density angular parameter $q$ & 1.0 \\
Angle between the equatorial plan and edge of the torus & $40^{\circ}$ \\
Ratio of the maximum to minimum radii of the torus & 20 \\
Viewing angle  &$30^{\circ}\,\,\rm{(type\,\,1)},70^{\circ}\,\,\rm{(type\,\,2)}$ \\
Disk spectrum  & Schartmann (2005) \\
Power-law index modifying the optical slope of the disk & -0.36\\
AGN fraction & 0.01, 0.1, 0.2, 0.3, 0.4, 0.5, 0.6, 0.7, 0.8, 0.9, 0.99 \\
Extinction law of polar dust & SMC \\
$E(B-V)$ of polar dust & 0.0, 0.01, 0.05, 0.1, 0.5, 1.0\\
Temperature of polar dust & 100 K \\
Emissivity of polar dust & 1.6 \\
\hline\\[-1.5ex]
\multicolumn{2}{c}{X-ray module} \\[0.5ex]
AGN photon index $\Gamma$ & 1.8 \\
$\alpha_{ox}$ & -1.9, -1.75, -1.6, -1.45, -1.3, -1.15, -1.0\\
Maximum deviation from the $\alpha_{ox}$--$L_\text{2500\AA}$ relation & 0.4 \\
LMXB photon index & 1.56 \\
HMXB photon index & 2.0 \\
\hline\\[-1.5ex]
Total number of models per redshift & 5,544,000 \\[0.5ex]
\hline
\label{table:cigale-input}
\end{tabular}
\end{table*}

\subsection{Fitting}
\label{subsec:33}
We run CIGALE for our final sample (347 sources) using the modules and parameters described in \cref{table:cigale-input}. The number of models computed per source by CIGALE is 5,544,000. We run it for two different types of miniJPAS photometry: \verb AUTO  and \verb PSFCOR .

Both photometries have their pros and cons. The \verb PSFCOR \ considers issues like point-spread function variation on the focal plane for different dates, biases on filters, and aperture correction, among others. However, its small aperture gives a value below the galaxy's expected total flux. \verb AUTO \ provides a closer value to the total flux, but it can be noisier due to a bigger aperture. To compare the SED fitting results of both photometric fluxes, we scale the J-spectra obtained from \verb PSFCOR  using the $r_\mathrm{mag}^\mathrm{AUTO}$ as the reference value.

\begin{figure}[!ht]
    \resizebox{\hsize}{!}{\includegraphics{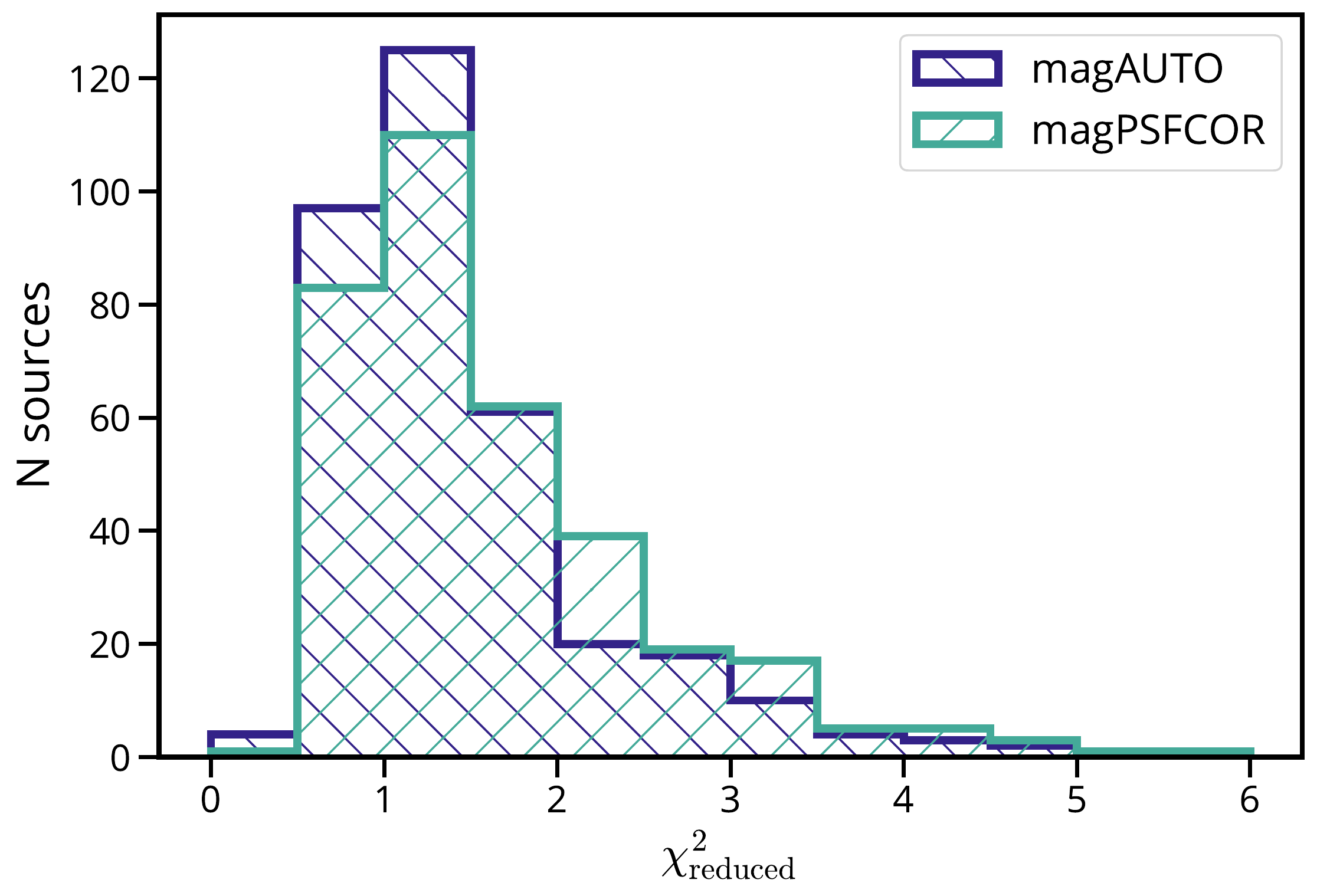}}
    \caption{Histograms of reduced $\chi^2$ for the SED fitting performed using two different sets of magnitudes extracted from miniJPAS catalogs, AUTO and PSFCOR (see \cshref{subsec:33}).}
    \label{fig:SED-redchi}
\end{figure}

\begin{figure*}[!t]
\centering
   \includegraphics[width=8.5cm]{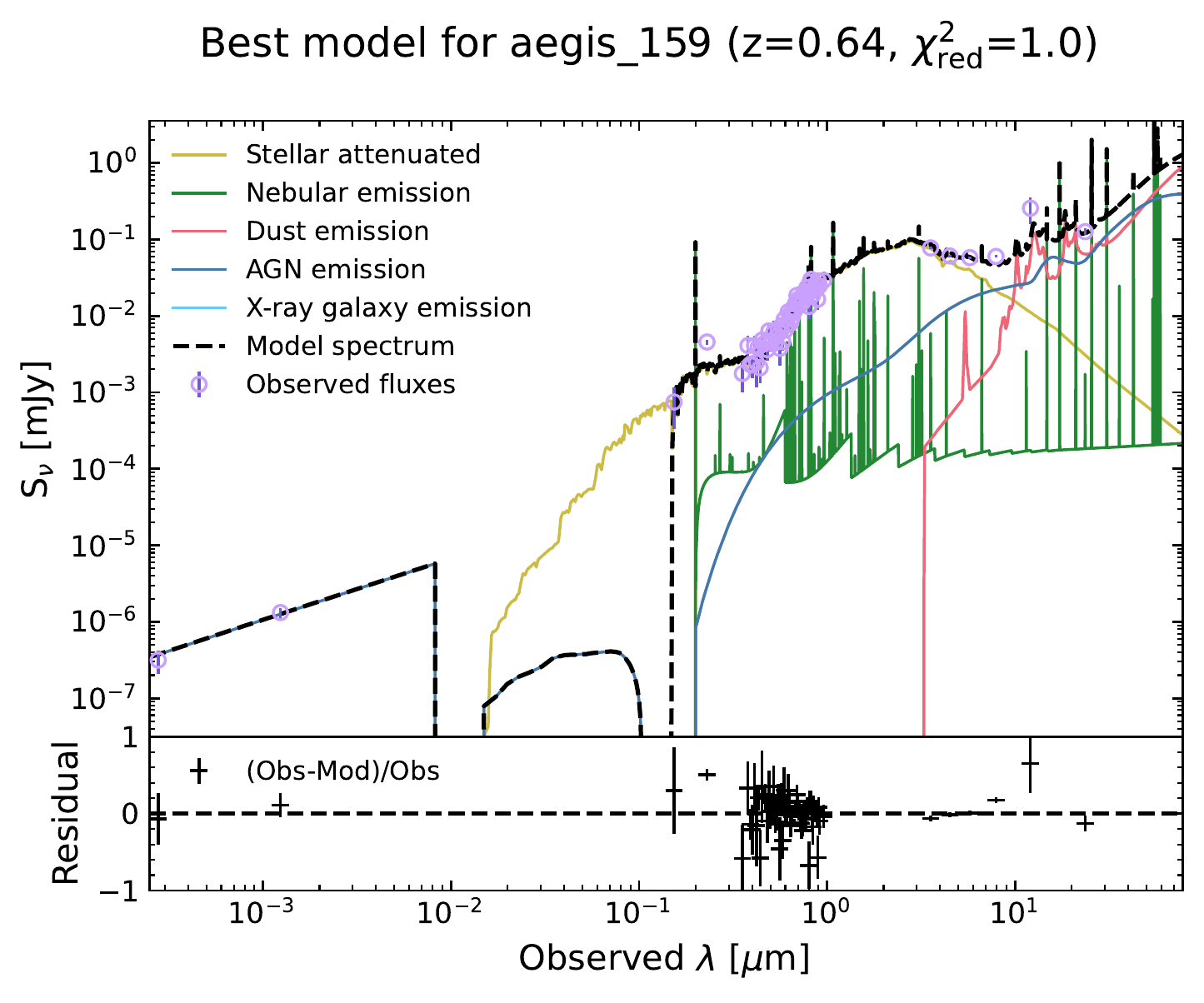}
   \includegraphics[width=8.5cm]{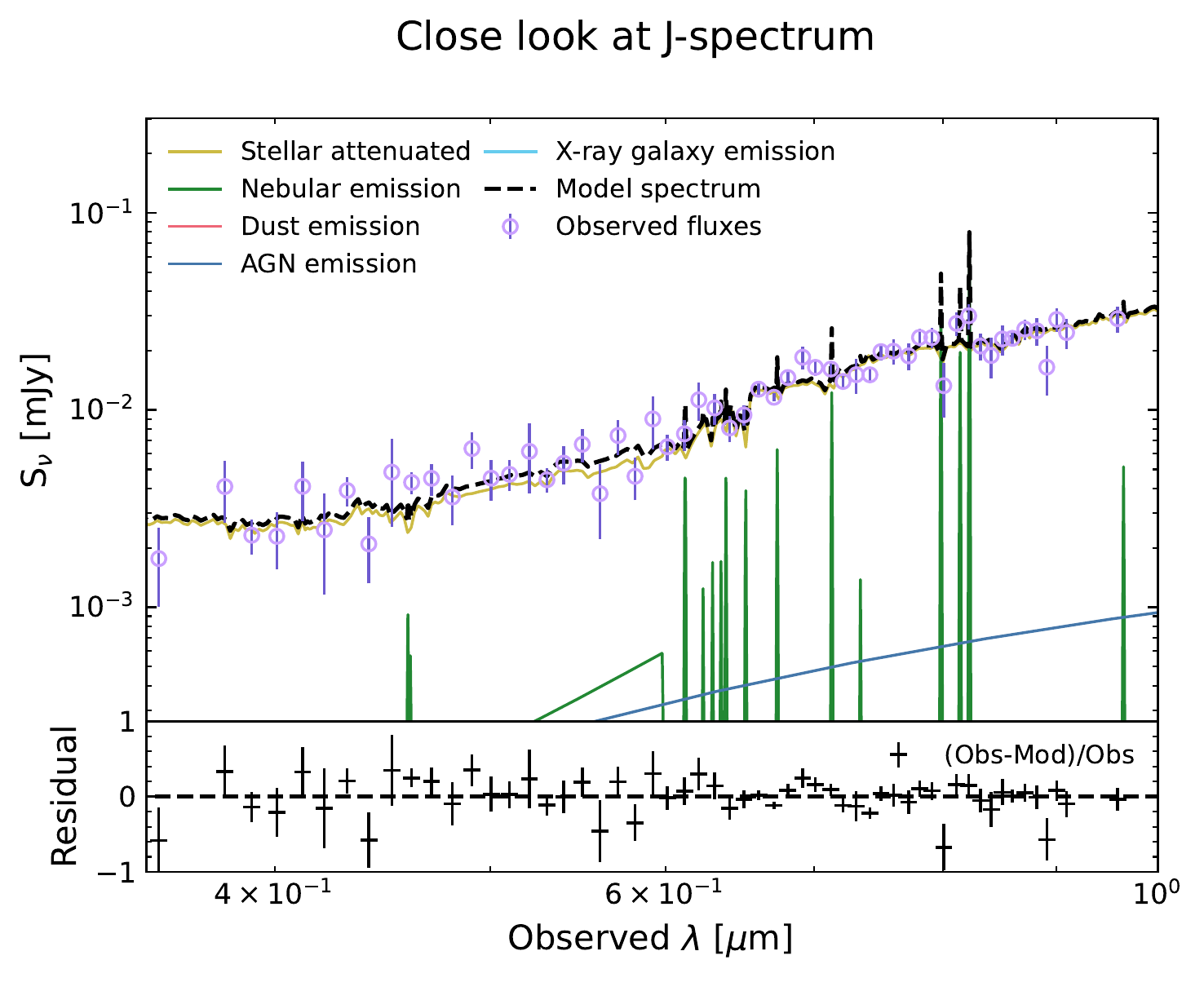}\\
   \includegraphics[width=8.5cm]{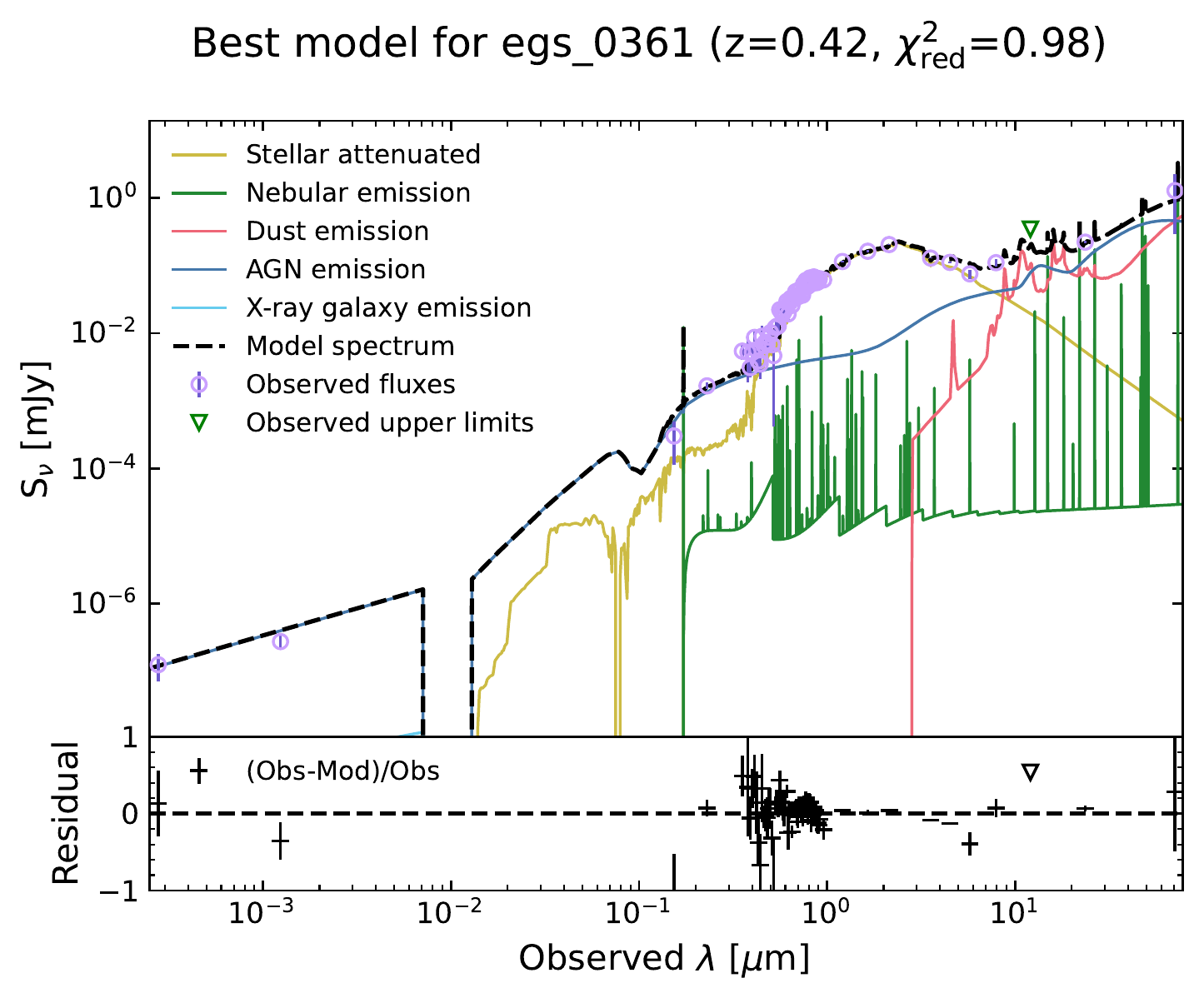}\includegraphics[width=8.5cm]{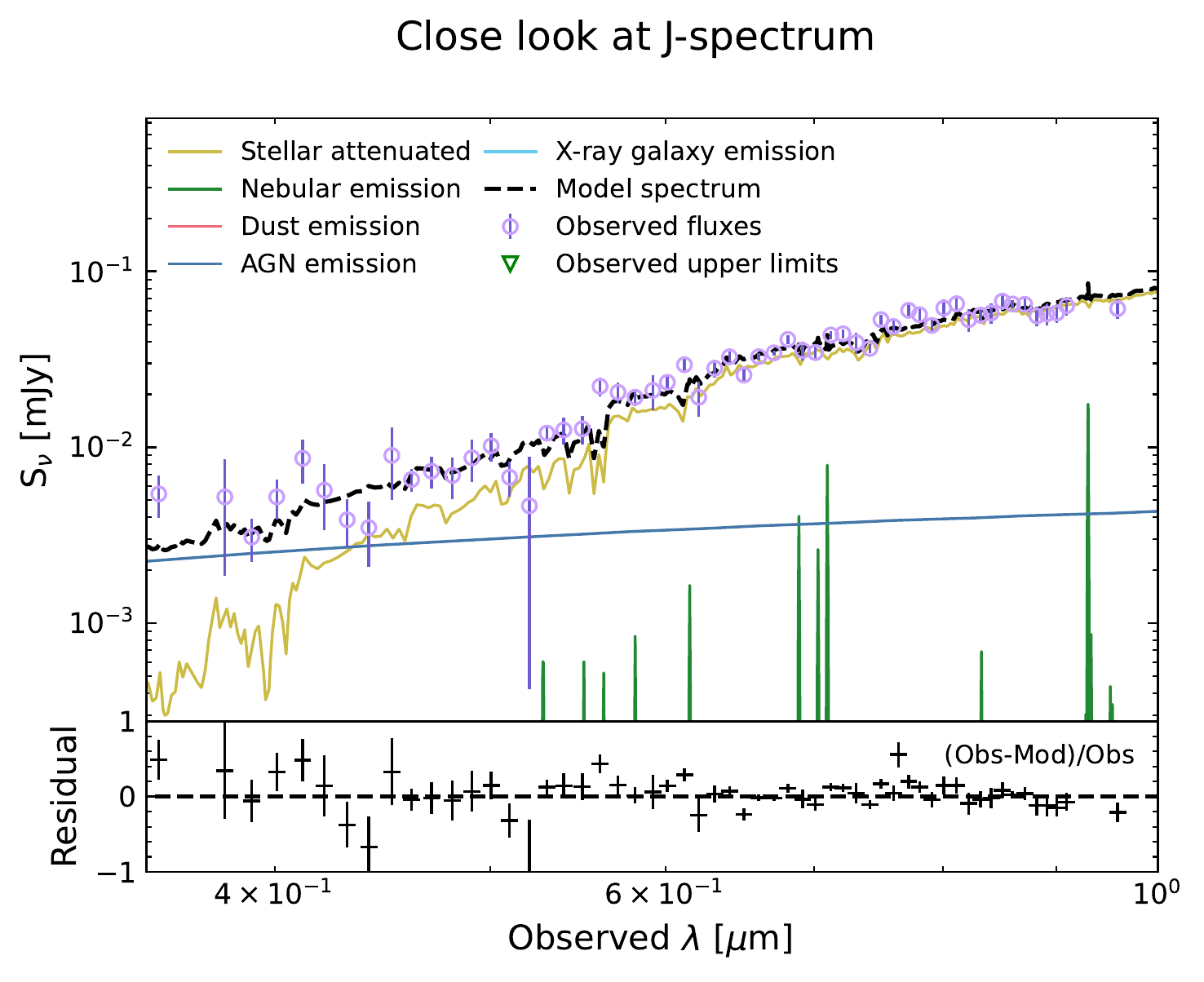}
   \caption{Examples of SED fitting using CIGALE and their residuals. Pink circles show the photometry for each band used. Green triangles are upper limits. The black dashed line is the composite model, and the color lines are the individual components of the composite model, as labelled. \textit{Left:} Full range of wavelengths, from X-ray to IR. \textit{Right:} Close up on optical miniJPAS J-spectrum.}
   \label{fig:SED-examples}
\end{figure*}

In Fig. \ref{fig:SED-redchi}, we show the distribution of $reduced\ \chi^2$ for the SED fitting with both types of photometries. Both distributions are similar, showing good fits, with a high number of sources near one and a decreasing tail beyond 3, being \verb AUTO \ the one with more sources near 1 (expected since \verb AUTO \ is noiser). For a deeper comparison between the results for both photometries, see \cref{appendix:phot}. Examples of SED fitting and its close up to the miniJPAS J-spectra are showed in \cref{fig:SED-examples}.

\begin{figure}
    \resizebox{\hsize}{!}{\includegraphics{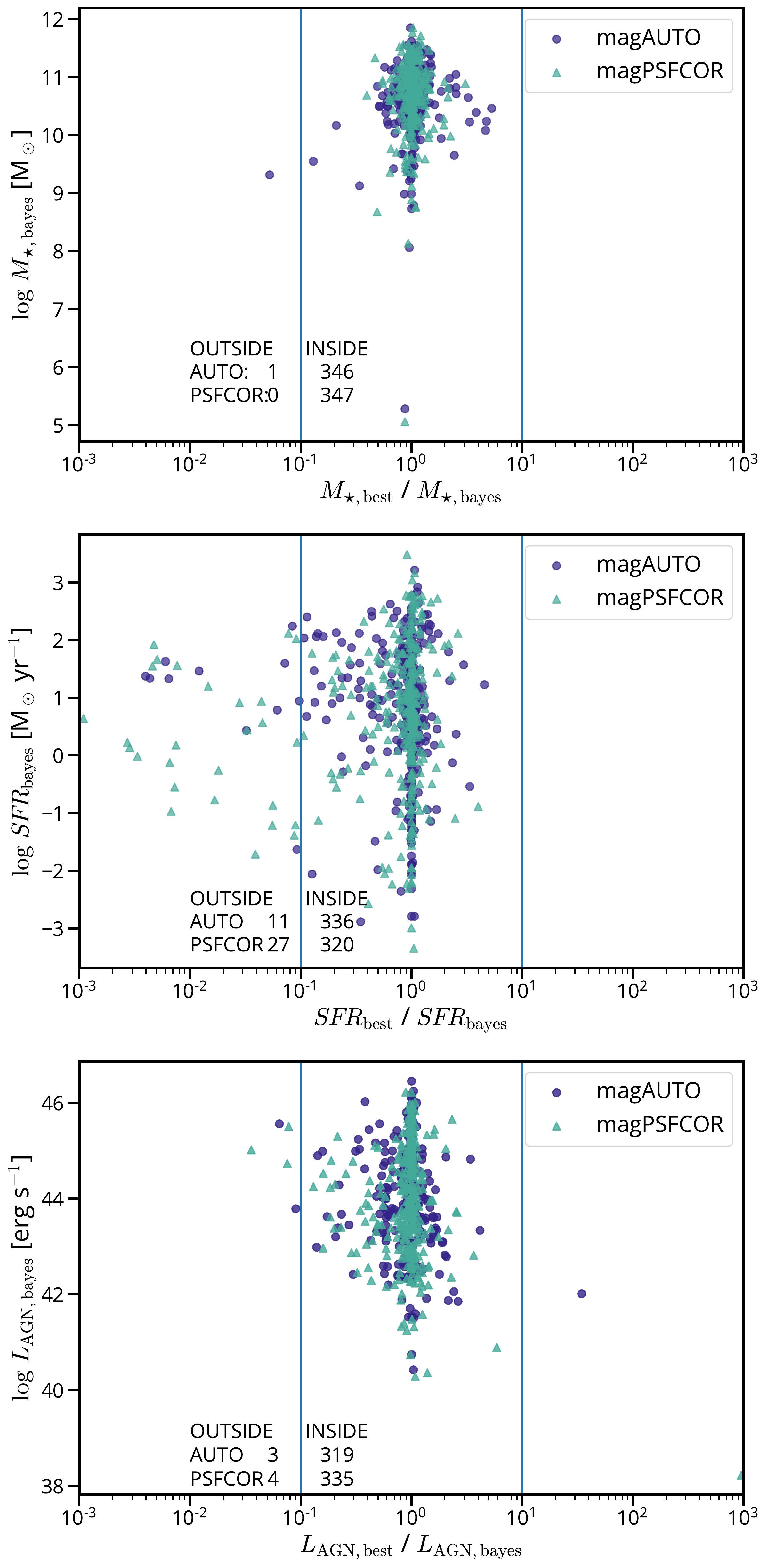}}
    \caption{Criteria used to exclude the sources with unreliable physical parameters. The bayesian values for $M_{\star}$ (upper panel), SFR (middle panel), and $L_{\rm AGN}$ (lower panel) are plotted against the ratio of best values over bayesian. The distribution is centered at 1. The solid vertical lines mark the limits of 0.1 and 10 adopted in this work (see Section 3.3 for details). The different colors show the parameters and ratios obtained assuming different magnitudes as input. The number of sources between and outside the limits is reported in the lower part of the plots.}
    \label{fig:SED-criteria}
\end{figure}

Since, in our analysis, it is necessary to obtain an estimation of the properties of the host galaxy and the AGN, we excluded the sources dominated by only one component (i.e., pure-AGN or pure-galaxy). \citet{2021A&A...653A..70M} and \citet{2021A&A...654A..93B} show that, for a given parameter, if there is a large difference between the \textit{best} value and the \textit{bayesian} value, the estimation of such parameter is not reliable. Following this idea, we can exclude the sources where the difference between parameters is bigger than one order of magnitude for $M_\star$, SFR, and $L_\mathrm{AGN}$. In other words, we can only keep sources with $0.1 \leq \frac{M_\mathrm{\star,best}}{M_\mathrm{\star,bayes}} \leq 10$,  $0.1 \leq \frac{L_\mathrm{AGN,best}}{L_\mathrm{AGN,bayes}} \leq 10$ and $0.1 \leq \frac{SFR_\mathrm{best}}{SFR_\mathrm{bayes}} \leq 10$. In \cref{fig:SED-criteria}, we show the reliability of our fits using these criteria. The parameters obtained with the different miniJPAS photometries show similar distributions, centered in 1 with small dispersion. The choice of 1 dex as limit was made to exclude all outliers that do not follow these distributions centered in 1. In Appendix \ref{appendix:phot}, we show no significant difference between the two photometries, besides that \verb AUTO \ has marginally smaller relative errors. From now on, we only present the results obtained using \verb AUTO . We also remove the sources where the AGN luminosity is close to 0. Finally, following these criteria, we remove 39 sources ($\sim$\,11\% of the sample). We are left with 308 sources with reliable measurements of the AGN and the host galaxy components.

\section{Data Analysis: optical spectra}
\label{sec:4}
The emission lines in AGN spectrum can provide information about their obscuration and the SMBH properties. Broad lines are observable for Type I, while narrow lines are present in both types. Typically, we can assume that the SMBH's gravitational field dominates the gas cloud motion in the broad-line region (BLR). Thus, the width of these lines is related to the virialized mass of the SMBH. A spectral fitting is necessary to obtain a good measure of the width, considering all other features typical of AGN. In this Section, we will discuss the spectra used, the fitting process, and the estimation of $M_\mathrm{BH}$.

\subsection{Fitting}
\label{subsec:41}
We described the spectra used in \cshref{subsec:24}, but in summary, at least one spectrum was available for 269 sources of our initial sample. We did a cut of a mean SNR\,$>$\,3 on these spectra. After we fit them, we obtained an acceptable FWHM of broad region lines of 113 AGN for our final sample. When more than one spectra were available for the same source, we selected the one with the highest SNR.

We used PyQSOFit \citep{2018ascl.soft09008G} to fit the continuum, iron emission, and emission lines of the AGN. We fitted the spectra with a combination of continuum and line emission. We used a polynomial for the stellar continuum and a power law for the AGN continuum, and we also included iron emission. In the case of narrow lines, we allow one gaussian with the same width for all the narrow features. Broad lines can be very complex because of the presence of asymmetries; in these cases, the width estimated by only one gaussian gives systematically larger widths \citep{2008ApJ...680..169S}. Due to this overestimation, we allow between one and three Gaussians depending on the line following \citet{2020ApJS..249...17R}. All the multi-component Gaussian fitted for narrow and broad emission lines are listed in \cref{table:lines}. In particular, we measured the FWHM of the broad component of H$\alpha$, H$\beta$, MgII, and CIV and the luminosity at 1350, 3000, and 5100 \AA . For the DEEP2 spectra that is not flux calibrated, we scaled them to the luminosity (1350, 3000, and 5100~\AA ) of the closest narrow-band filter from miniJPAS photometry.

\begin{table}
\caption{Emission lines fitted. Central wavelengths are at rest-frame.}
\label{table:lines}
    \centering          
    \begin{tabular}{l c c c }
    \hline\hline       
    Line & Wavelength [\AA ] & Type & Gaussians\\
    \hline
    H$\alpha$ & 6564.61 & Broad  & 3\\
    H$\alpha$ & 6564.61 & Narrow & 1\\
    $[$NII]       & 6549.85 & Narrow & 1\\
    $[$NII]       & 6585.28 & Narrow & 1\\
    $[$SII]       & 6718.29 & Narrow & 1\\
    $[$SII]       & 6732.67 & Narrow & 1\\
    H$\beta$  & 4862.68 & Broad  & 3\\
    H$\beta$  & 4862.68 & Narrow & 1\\
    $[$OIII]      & 4960.30 & Narrow & 1\\
    $[$OIII]      & 5008.24 & Narrow & 1\\
    $[$OII]       & 3728.48 & Narrow & 1\\
    $[$SII]       & 6732.67 & Narrow & 1\\
    MgII      & 2798.75 & Broad  & 2\\
    MgII      & 2798.75 & Narrow & 1\\
    CIII]      & 1908.73 & Broad  & 1\\
    CIV       & 1549.06 & Broad  & 3\\
    CIV       & 1549.06 & Narrow & 1\\
    Ly$\alpha$& 1215.67 & Broad  & 1\\
    Ly$\alpha$& 1215.67 & Narrow & 1\\
    \hline
    \end{tabular}
\end{table}

We set limits for the width of the lines fitted. To distinguish between broad and narrow lines, we used a value of 1,000~km~s$^{-1}$. The upper limit for the broad lines was $10\,000$~km~s$^{-1}$. These limits come from the width bimodal distribution shown for X-ray-selected AGN lines \citep{2016MNRAS.457..110M}.

\subsection{BH mass estimation}
\label{subsec:42}

Although $\sigma$ seems to correlate better with the masses calculated from the reverberation method, we use FWHM instead. The main criterion for this choice is that $\sigma$ is too sensitive to noise on the wings of the emission lines\citep{2012ApJ...753..125S}. To estimate the error in the FWHM, PyQSOFit uses a Monte-Carlo approach to fit random mock spectra and obtain the uncertainties considering the flux errors and systematic errors for multiple decomposing components. 

Several works calibrate the virial relation from a single-epoch spectrum \citep[see][for a compilation]{2012ApJ...753..125S}. Different lines have slightly different calibrations. For our $M_\mathrm{BH}$ estimation, we used the coefficients from \citet{2011ApJ...742...93A} for the Balmer lines (H$\alpha$ and H$\beta$); \citet{2009ApJ...699..800V} for MgII; and \citet{2006ApJ...641..689V} for CIV. We used the following equation to estimate the black hole masses, with an overview of the coefficients in \cref{table:coefs}.

\begin{equation}
    \log \left( \frac{M_\mathrm{BH,vir}}{M_\odot}\right) = a + b\log \left( \frac{L}{10^{44} \mathrm{ erg s}^{-1}}\right) + c \log \left( \frac{\mathrm{FWHM}}{\mathrm{km s}^{-1}} \right)
    \label{eq:Virial}
\end{equation}

\begin{table}
\caption{Coefficients used for different emission lines. Extracted from \citet{2012ApJ...753..125S}.}
\label{table:coefs}
    \centering          
    \begin{tabular}{l c c c c }     
    \hline\hline       
    FWHM & Luminosity measure at & a & b & c \\
    \hline
    H$\alpha$&5100~\AA &0.774&0.520&2.06\\
    H$\beta$&5100~\AA  &0.895&0.520&2.00\\
    MgII&3000~\AA      &0.860&0.500&2.00\\
    CIV&1350~\AA       &0.660&0.530&2.00\\
    \hline
    \end{tabular}
\end{table}

\begin{figure*}[!t]
\centering
   \includegraphics[width=8.5cm]{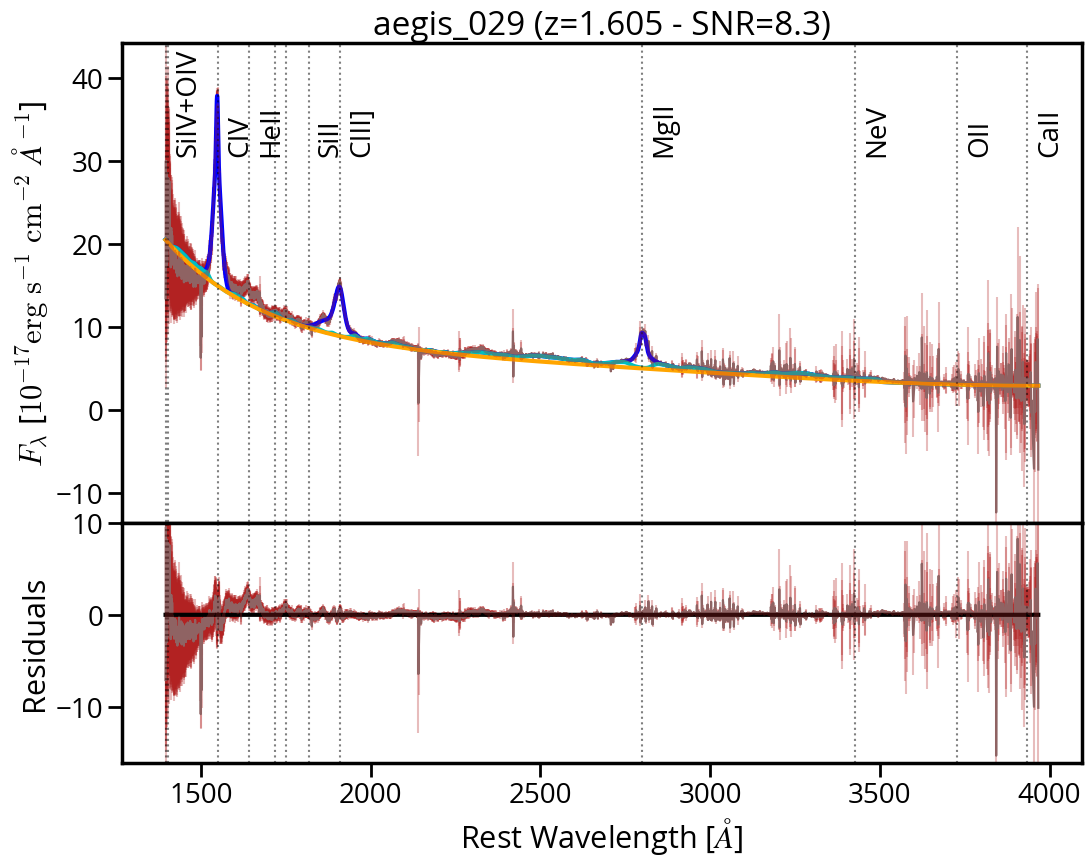}
   \includegraphics[width=8.5cm]{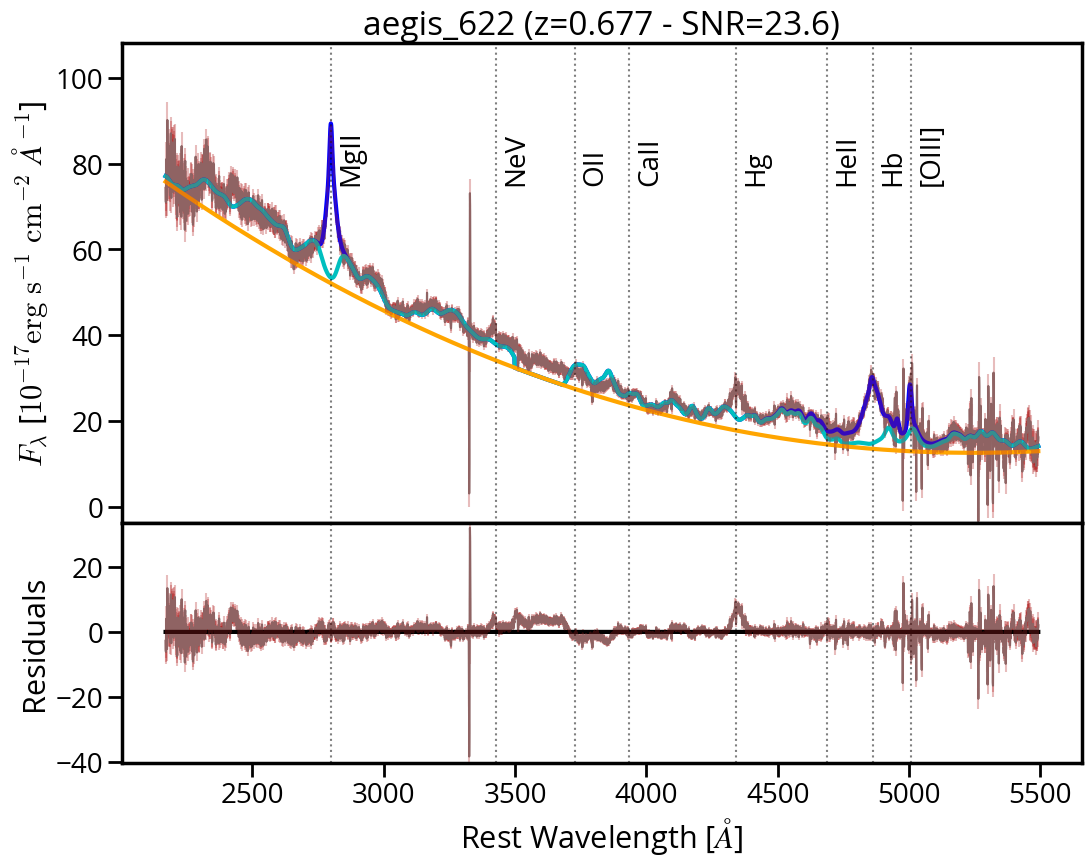}\\
   \includegraphics[width=8.5cm]{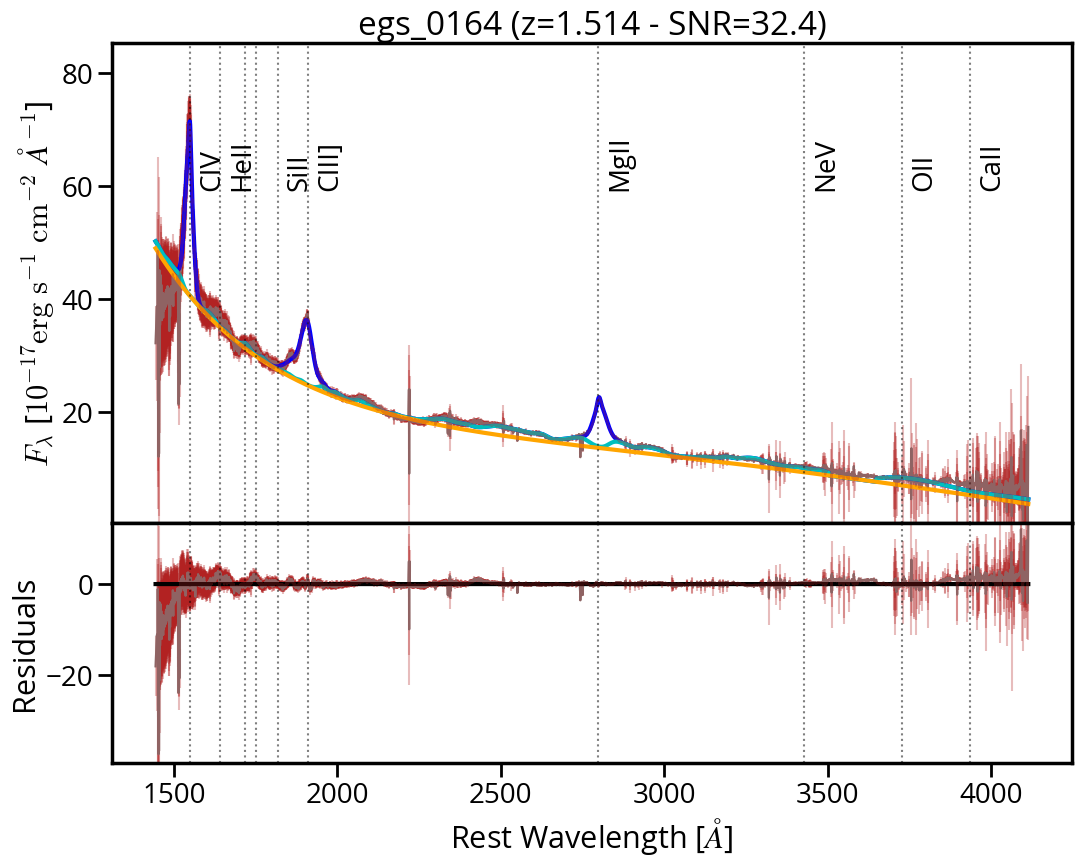}
   \includegraphics[width=8.5cm]{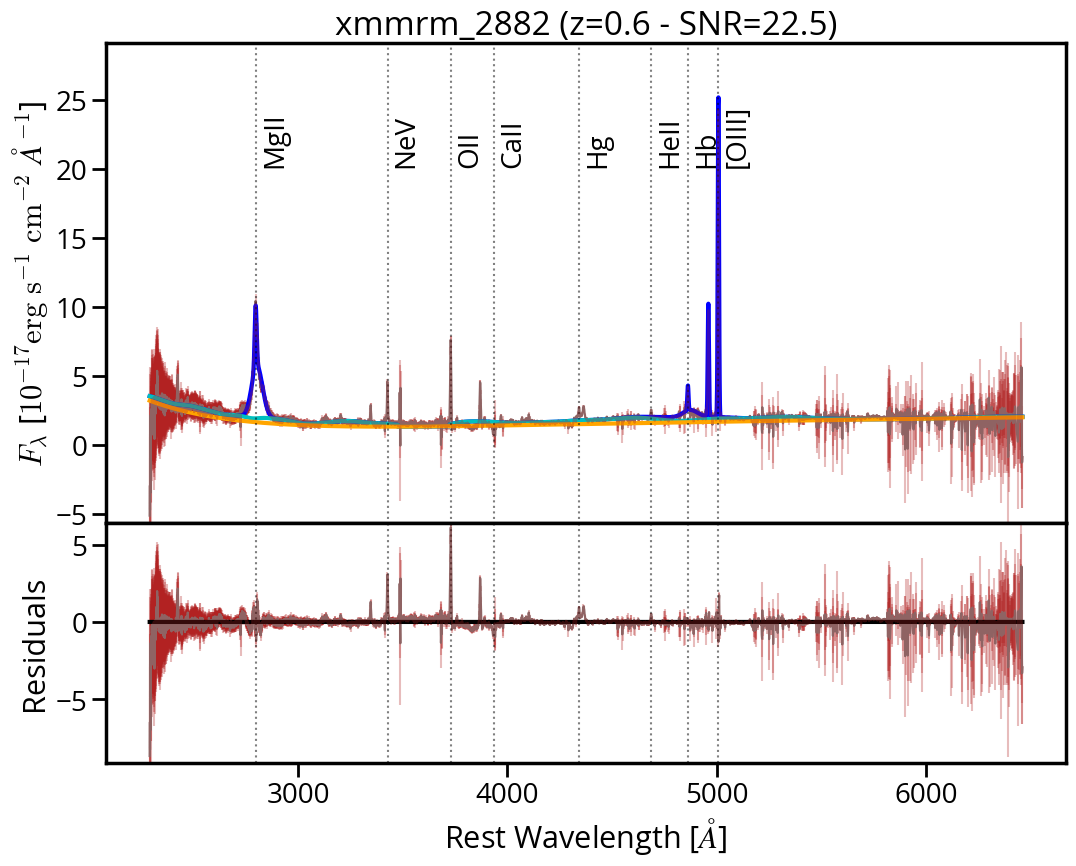}
   \caption{Examples of spectral fitting using PyQSOFit, and their residuals. On grey the spectra observed, at rest-frame wavelength. Continuum was fitted as a polynomial, showed in orange. On cyan, the Fe emission. On blue, the emission lines fitted in the process to obtain FWHM of the BLR.}
   \label{fig:Spectra-examples}
\end{figure*}

To obtain $M_\mathrm{BH}$ uncertainties, we propagate the FWHM and luminosity uncertainties in the \cref{eq:Virial}, including the dispersion of the original fit where the coefficients $a$, $b$, and $c$ were calculated. Similar to the spectra selection, to estimate $M_\mathrm{BH}$ when more than one line was present, we selected the one with higher SNR (if the corresponding fit was acceptable). Examples of spectral fits are shown in \cref{fig:Spectra-examples}. Finally, we obtained black hole masses for 113 sources in our sample with reliable SED fitting. Sixty of these masses were obtained using SDSS data, 43 from MMT and 10 from DEEP2.

\begin{figure}
    \resizebox{\hsize}{!}{\includegraphics{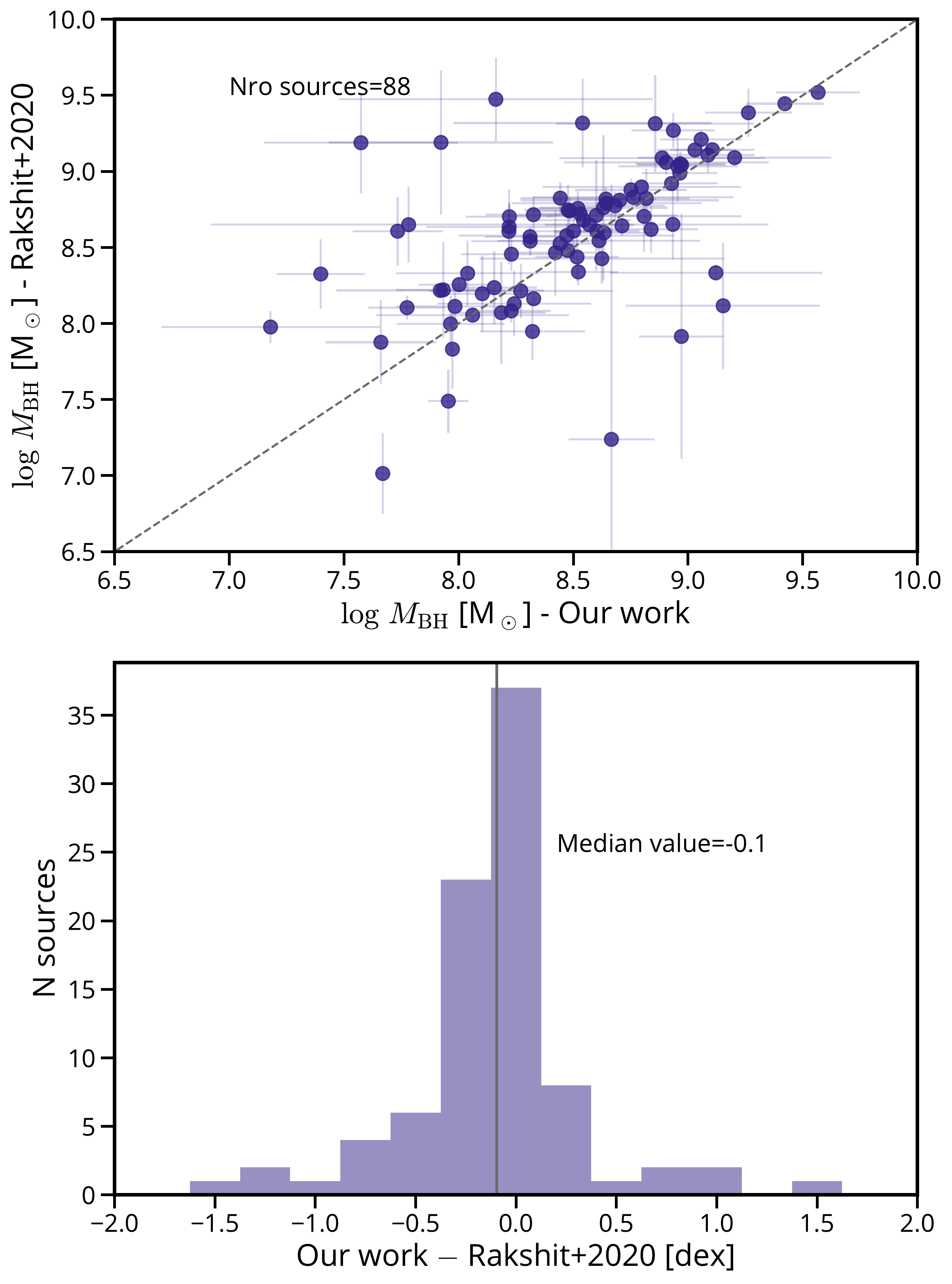}}
    \caption{{\it Upper panel}: Comparison between our estimation of the black hole masses and the ones estimated by \citet{2020ApJS..249...17R} for the 88 sources in common. {\it Bottom panel}: Histogram of the logarithmic difference of the BH masses. }
    \label{fig:BH_Rakshit}
\end{figure}

These masses are similar to those obtained by \citet{2020ApJS..249...17R}, as shown in \cref{fig:BH_Rakshit}. We do not find a systematic shift between our masses and theirs. The difference for some sources may be related to the difference in the fitted spectra; while they use the best SDSS spectra, we use SDSS stacked spectra or a different epoch MMT spectra, if the SNR was higher than SDSS. Because our work also considers stacking and adds spectra from other telescopes, we obtained BH masses for more sources in our sample. Another way to obtain spectra independent $M_\mathrm{BH}$ is by measuring the broad line directly from the J-spectra. \citet{2022A&A...660A..95C} explored this possibility with promising results compared with \citet{2020ApJS..249...17R}. While this work can be applied to extensive areas in the future of JPAS, it is currently limited to sources brighter than $r<21$ and $M_\mathrm{BH}>10^8$~M$_\odot$.

\section{Physical properties of AGN and host galaxies}
\label{sec:5}

Until now, we have recovered reliable values of properties for 308 AGN and their host galaxies from SED fitting. We also estimate $M_\mathrm{BH}$ for a subsample of 113 sources. In this Section, we show the distributions of these properties and the derivation of additional parameters that depend on them.

\subsection{Distributions of physical properties}
\label{subsec:51}
\begin{figure}
    \resizebox{\hsize}{!}{\includegraphics{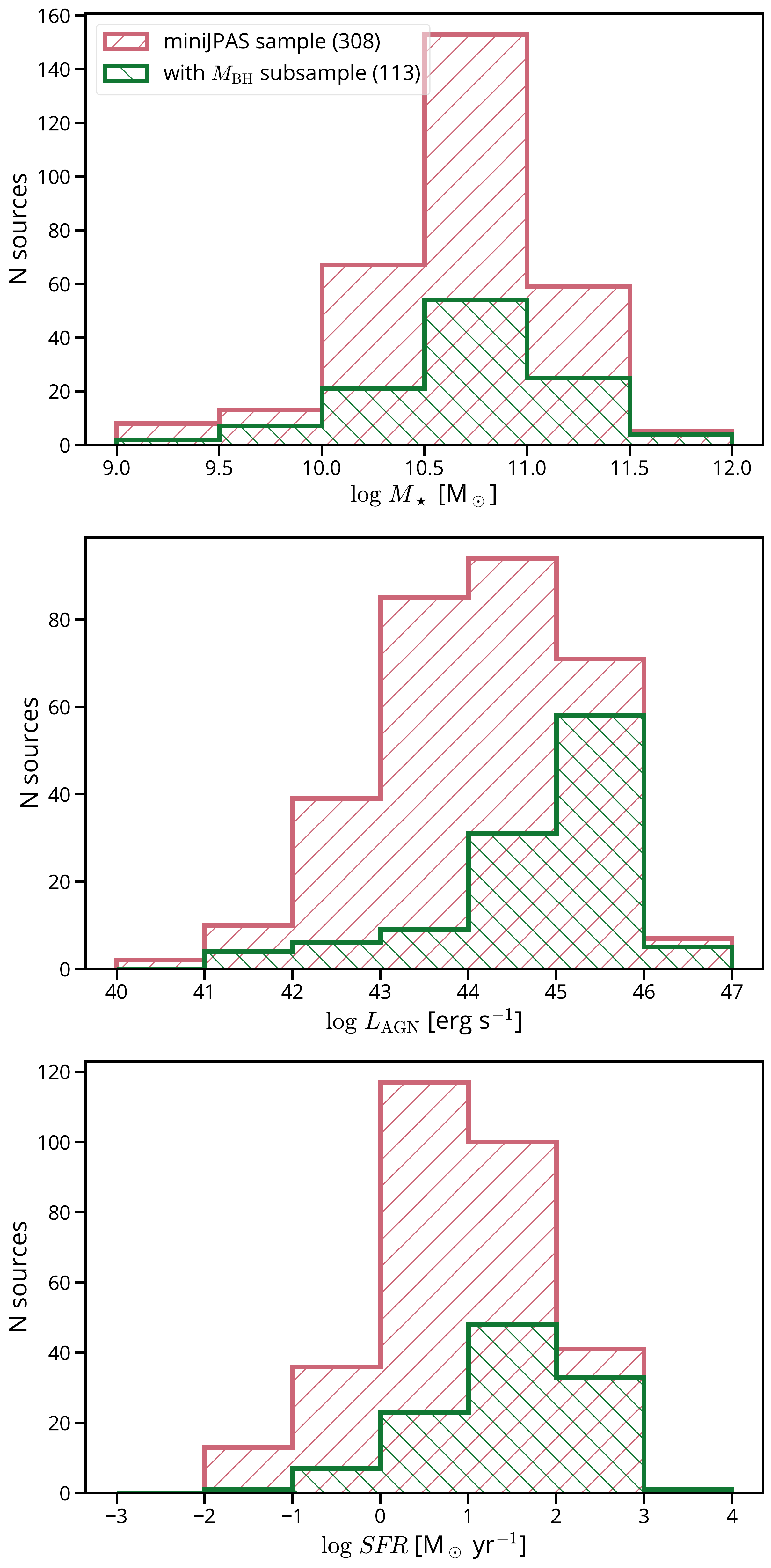}}
    \caption{Histograms of the physical parameters estimated with the SED fitting ($M_\star$, $L_\mathrm{AGN}$ and SFR). In red is the full sample while in green is the subsample with an estimation of $M_\mathrm{BH}$ using the corresponding spectra.}
    \label{fig:SED-Hist}
\end{figure}

In \cref{fig:SED-Hist}, we show the distributions of $M_\star$, SFR, and $L_\mathrm{AGN}$ for the entire sample of 308 miniJPAS sources for which these parameters have been derived (red) and the subsample of 113 sources for which we also have a reliable measurement of  $M_\mathrm{BH}$ (green). The values for individual sources can be found in \cref{t_prop}. While the distribution in $M_\star$ is very similar for the two samples, the subsample with measured BH masses is biased towards higher $L_\mathrm{AGN}$ and SFR.

To quantify the difference, we performed a Kolmogorov-Smirnov test to check the null hypothesis that the subsample was drawn from the same probability distribution of the larger sample. We obtained a \textit{p-value} smaller than 1\% for $L_\mathrm{AGN}$ and SFR. While we cannot discard the null hypothesis for the $M_\star$ (\textit{p-value} $\sim$\,0.88), we can say our subsample shows a bias towards high values of $L_\mathrm{AGN}$ and SFR. The origin of this bias could be ascribed to the fact that it is easier to obtain well defined spectra for more luminous sources, which in turn are powered, on average, by more massive black holes. 

\begin{table*}
\caption{\label{t_prop}Physical properties of the galaxies in the entire sample of 308 galaxies.}
\centering
\begin{tabular}{lcccccc}
\hline\hline
id & $\log$ $M_\star$ &       SFR &   $\log$   $M_\mathrm{BH}$ & $\log$ $L_\mathrm{AGN}$ &  BHAR &       $\log \lambda_\mathrm{Edd}$\\
&[M$_\odot$]&[M$_\odot$ yr$^{-1}$]&[M$_\odot$]&[erg s$^{-1}$]&[M$_\odot$ yr$^{-1}$]&\\
\hline\\
aegis\_019 &   10.06 $\pm$ 0.26 &   6.73 $\pm$   3.98 &     -            & 43.60 $\pm$ 0.55 &     - &     - \\
aegis\_021 &   11.37 $\pm$ 0.19 & 134.60 $\pm$ 174.72 &     -            & 46.25 $\pm$ 0.08 &     - &     - \\
aegis\_022 &   11.10 $\pm$ 0.22 & 124.05 $\pm$  50.57 &     -            & 45.22 $\pm$ 0.05 &     - &     - \\
aegis\_029 &   10.39 $\pm$ 1.28 &  37.11 $\pm$  29.93 &  8.48 $\pm$ 0.18 & 45.80 $\pm$ 0.05 &  1.11 & -0.80 \\
aegis\_032 &   10.93 $\pm$ 0.22 &   5.36 $\pm$   0.72 &     -            & 45.26 $\pm$ 0.18 &     - &     - \\
aegis\_035 &   10.31 $\pm$ 0.12 &   1.21 $\pm$   0.27 &  7.77 $\pm$ 0.38 & 43.10 $\pm$ 0.41 &  0.05 & -2.78 \\
aegis\_036 &   10.75 $\pm$ 0.25 &   7.63 $\pm$   4.55 &  8.48 $\pm$ 0.01 & 44.55 $\pm$ 0.23 &  0.66 & -2.04 \\
aegis\_037 &   10.26 $\pm$ 0.26 &  72.46 $\pm$   7.58 &  7.78 $\pm$ 0.17 & 45.07 $\pm$ 0.06 &  0.21 & -0.82 \\
...&...&...&...&...&...&...\\
\hline
\end{tabular}
\tablefoot{Full table available online in digital format.}
\end{table*}

\subsection{Eddington ratios and proxies for the accretion rates}
\label{subsec:52}
\begin{figure}[h!]
    \resizebox{\hsize}{!}{\includegraphics{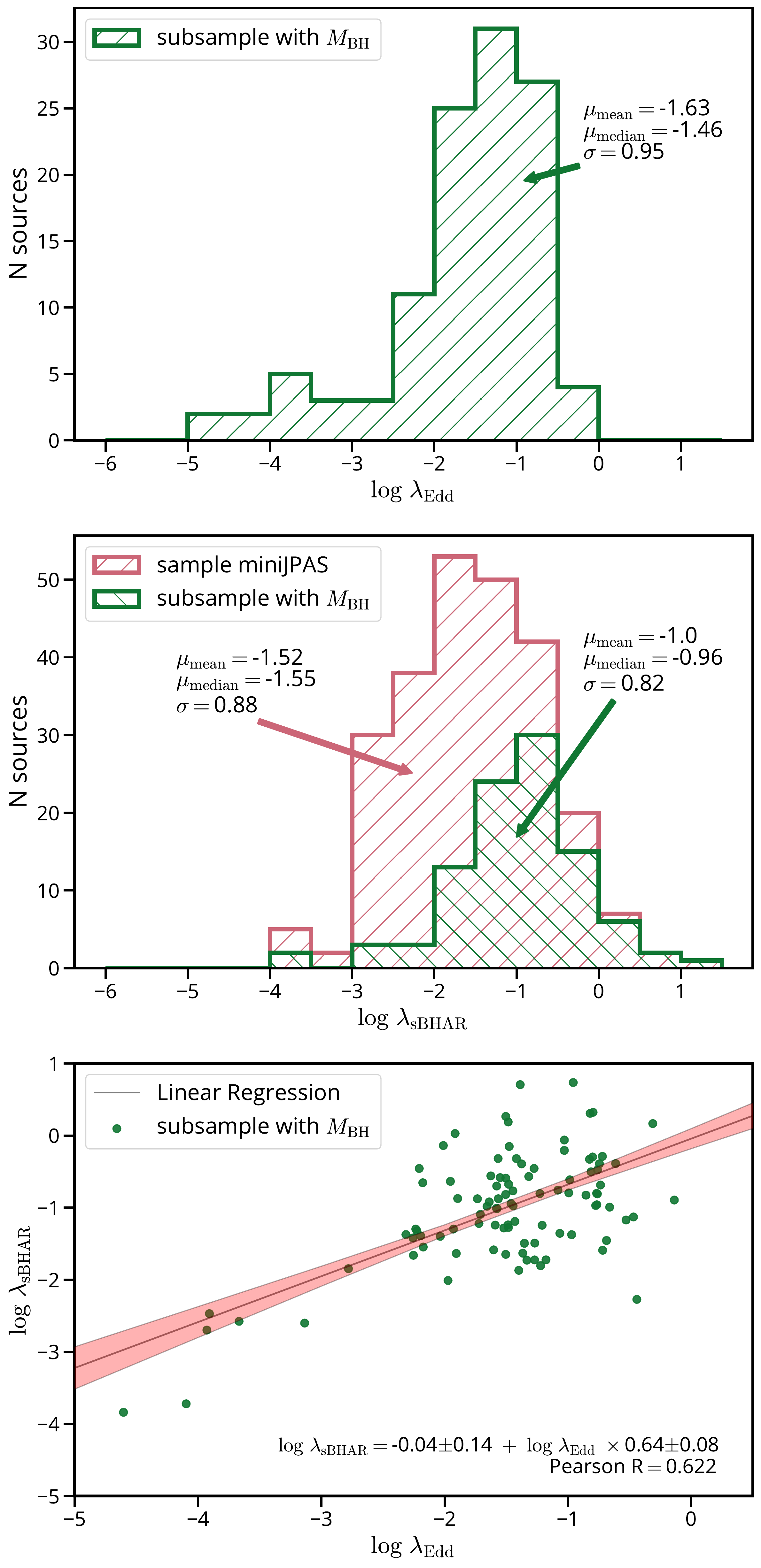}}
    \caption{Distribution of accretion rates derived in this work. {\it Upper panel:} Histogram of Eddington accretion rate ($\lambda_\mathrm{Edd}$) for the sources with measured $M_\mathrm{BH}$. {\it Middle panel:} Histogram of the specific accretion rate ($\lambda_\mathrm{sBHAR}$,   typically used as proxy of $\lambda_\mathrm{Edd}$), for both the sample with measured BH masses and the full miniJPAS sample. A difference in the shape of  distributions can be seen for the subsample with measured BH masses by comparing the green histograms in the upper and middle panels.
    {\it Lower panel:} Plot of the relation between $\lambda_\mathrm{sBHAR}$--$\lambda_\mathrm{Edd}$. While they are defined to be equal, we found a poor correlation value and a systematic shift from the 1:1 relation in their linear regression (plotted as grey with 1-$\sigma$ uncertainties in pink).}
    \label{fig:Acrettion_ratios}
\end{figure}

We derived estimates of the Eddington ratios, which is a fundamental parameter for constraining BH cosmological evolution (\cshref{sec:1}), for the subsample of 113 sources with BH mass estimates based on BLR widths. The distribution of $\lambda_\mathrm{Edd}$ for this subsample is shown in the upper panel of Figure  \ref{fig:Acrettion_ratios}. This distribution shows a clear peak at around $\lambda_\mathrm{Edd}$\,$\sim$\,0.1, with a steep fall off at larger Eddington values, and a slightly less abrupt one at lower Eddington ratios. 

The Eddington ratio distribution of our sample is significantly different from other works \citep{2009ApJ...699..800V,2012ApJ...761..143N,2012MNRAS.425..623L}. In particular, we compared with \citet{2012MNRAS.425..623L} because both samples were X-ray selected, and physical properties were estimated with similar methods (SED and broad-line fitting). The Eddington ratio distributions have a difference of 0.5 (0.7) dex for median (mean) values. The difference is likely due to the combination of selection effects and estimation of $L_\mathrm{AGN}$ in each study. Both samples are similar in $L_\mathrm{X}$, but their optical spectra are deeper, reaching higher magnitudes than our work. Second, their estimation of $L_\mathrm{AGN}$ comes from the bolometric AGN emission from 1 $\mu$m to 200 keV; in our case, $L_\mathrm{AGN}$ only corresponds to the integrated accretion disk luminosity (\verb agn.accretion_power \ in CIGALE). In our sample, we registered a difference in 0.3 dex comparing our $L_\mathrm{AGN}$ with the higher values of bolometric AGN luminosity calculated from CIGALE.

For completeness, following other works in the literature \citep[e.g.,][]{2018MNRAS.474.1225A}, we also compute distributions using alternative measurements of the Eddington ratios, namely $\lambda_\mathrm{sBHAR}$, defined as: 
\begin{equation*}
    \lambda_\mathrm{sBHAR} = \frac{k_\mathrm{bol}~L_\mathrm{X}}{1.3\times10^{38}~\mathrm{erg~s}^{-1}\times 0.002\frac{M_\star}{\mathrm{M}_\odot}}
\end{equation*}
For comparison, we adopt $k_\mathrm{bol}$ as 25. We note that in this case we can include all sources from our AGN sample as $\lambda_\mathrm{sBHAR}$ does not rely on independent BH mass measurements. This rate is defined to be $\lambda_\mathrm{sBHAR}$\,$\approx$\,$\lambda_\mathrm{Edd}$ using the strong hypothesis that the mass of the SMBH scales directly with the total stellar mass of the host galaxy, as $M_\mathrm{BH}\sim 0.002~M_\star$ \citep{2003ApJ...589L..21M,2018MNRAS.474.1225A}.

We show the $\lambda_\mathrm{sBHAR}$ for the entire miniJPAS X-ray sample (red histogram) and also for the subsample with measured BH mass (green histogram) in bottom panel of \cref{fig:Acrettion_ratios}. We observe a difference between the $\lambda_\mathrm{Edd}$ and its proxy $\lambda_\mathrm{sBHAR}$ with the latter being $\sim$\,0.6 dex larger. When we compare these distributions with a Kolmogorov-Smirnov test, we obtained a \textit{p-value} lower than 1\%, further suggesting that these distributions are significantly different, shedding some doubts on the actual suitability of $\lambda_\mathrm{sBHAR}$ in constraining BH accretion rate models. 

While these two rates are defined to be similar, the difference is outstanding. The strong assumption used to acquire an $\lambda_\mathrm{Edd}$ without a measured BH mass can explain this difference since our sample shows a strong scatter for the $M_\mathrm{BH}$--$M_\star$ relation. In \cref{sec:6}, we discuss the nature of the scaling relation for our selection and its possible evolution across cosmic times.

\subsection{AGN properties}
\label{subsec:53}
In \cshref{sec:1}, we mentioned the importance of obtaining BHAR distributions. This rate in physical units  ($\Dot{M}_\mathrm{BH}$) can be derived starting from the AGN accretion luminosity ($L_\mathrm{AGN}$). This is possible by assuming a proportionality between $L_\mathrm{AGN} \propto \Dot{M}_\mathrm{BH}/\Dot{M}_\mathrm{Edd}$, for all Eddington ratios ($\lambda_\mathrm{Edd}$). 
Given that we have an estimate of $M_\mathrm{BH}$, we followed \citet{2004MNRAS.353.1035M} and adopted a broken power-law, connecting the low accretion rate (radiatively inefficient) regime with the high accretion rate one. The break is given at $\lambda_\mathrm{crit} = 3\times 10^{-2}$ \citep{2008MmSAI..79.1310M}. Imposing continuity at $\lambda_\mathrm{crit}$, yields the following equation:

\begin{equation}
    \label{eq:dotmass}
    \Dot{M}_\mathrm{BH} = \left.
    \begin{cases}
		\mathlarger{ \frac{L_\mathrm{AGN}}{\eta c^2}}  & \mbox{if } \lambda_\mathrm{Edd} \geq \lambda_\mathrm{crit} \\
		&\\
		\mathlarger{\frac{\sqrt{\lambda_\mathrm{crit} L_\mathrm{AGN} L_\mathrm{Edd}}}{\eta c^2}} & \mbox{if } \lambda_\mathrm{Edd} < \lambda_\mathrm{crit}
    \end{cases}
    \right .
\end{equation}

where $\eta$ is the radiative efficiency, assumed as $0.1$. 

\begin{figure}
    \resizebox{\hsize}{!}{\includegraphics{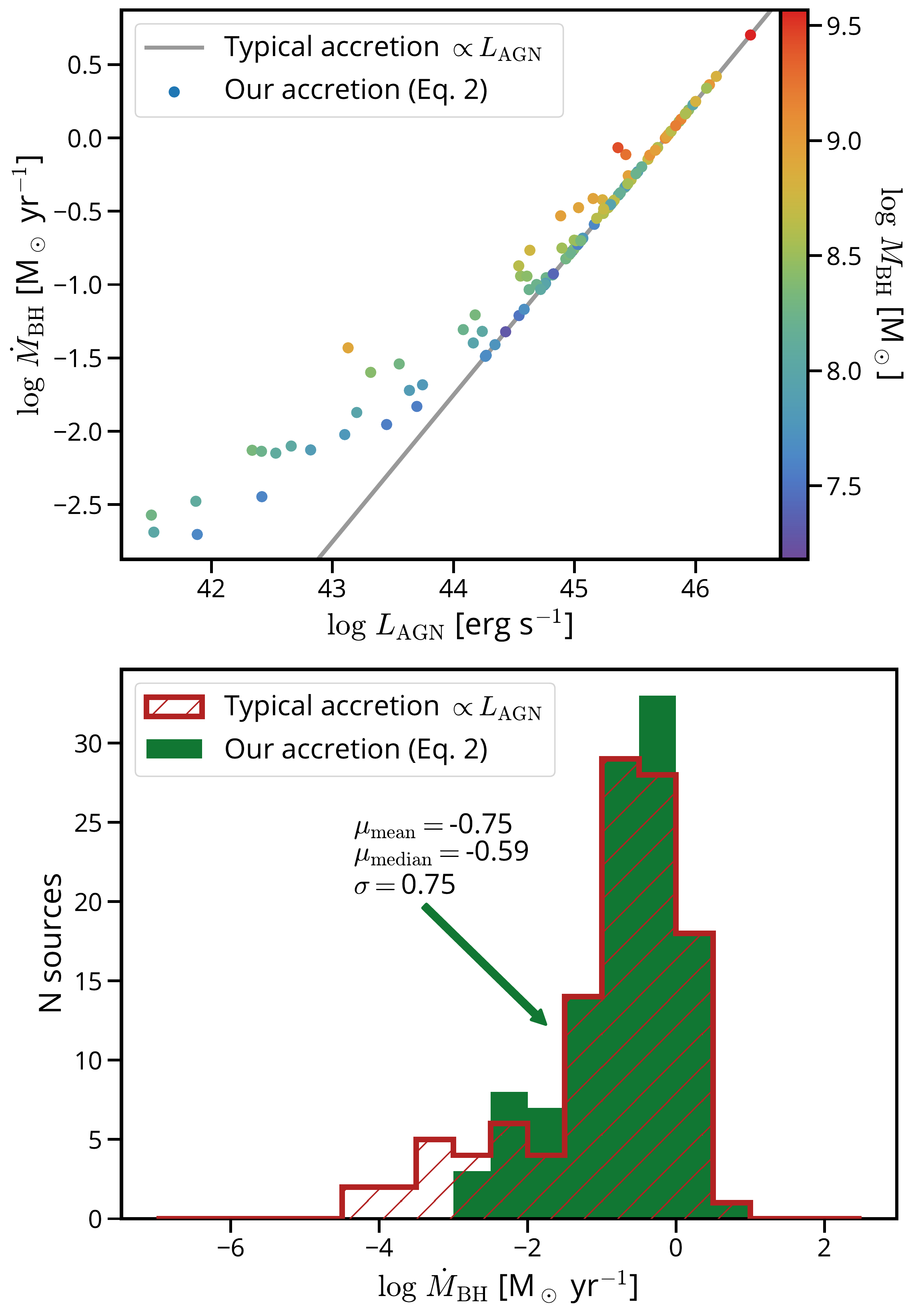}}
    \caption{\textit{Upper panel:} Accretion rate depending on $L_\mathrm{AGN}$. The solid line is the typical accretion rate assuming a linear dependence with $L_\mathrm{AGN}$, and it is equal for our sources with $\lambda > \lambda_{crit}$. For the rest of the sources, the dispersion comes from the dependence with $M_\mathrm{BH}$, which is color-coded. \textit{Lower panel:} Histogram of the estimated BH accretion rates.}
    \label{fig:Acrettion}
\end{figure}

We applied \cref{eq:dotmass} to derive the BHAR in physical units for our sources with measured $M_\mathrm{BH}$. \cref{fig:Acrettion} shows $\Dot{M}_\mathrm{BH}$ as a function of the AGN luminosity, color-coded by BH mass and the distribution of $\Dot{M}_\mathrm{BH}$. The shape of the distribution for higher values is, as expected, similar to the distribution of $\lambda_\mathrm{Edd}$ (see \cref{fig:Acrettion_ratios}). For lower values, it is less steep than $\lambda_\mathrm{Edd}$, as a consequence of the change in the accretion regime. It is also less steep than a distribution derived from $\Dot{M}_\mathrm{BH}\propto L_\mathrm{AGN}$, with a $\sigma$\,$=$\,0.75 instead of $\sigma$\,$=$\,0.95. With \cref{eq:dotmass}, we recovered the change in the accretion mode, obtaining higher values of $\Dot{M}_\mathrm{BH}$ than the typical accretion (as seen in the both panels of \cref{fig:Acrettion}). Our $\Dot{M}_\mathrm{BH}$ also shows similar correlations with SFR as \citet{2015MNRAS.449..373D} and \citet{2018A&A...618A..31M}, even with the two regimes.

\section{Stellar and BH mass evolution}
\label{sec:6}

\begin{figure*}[!t]
\centering
   \includegraphics[width=17cm]{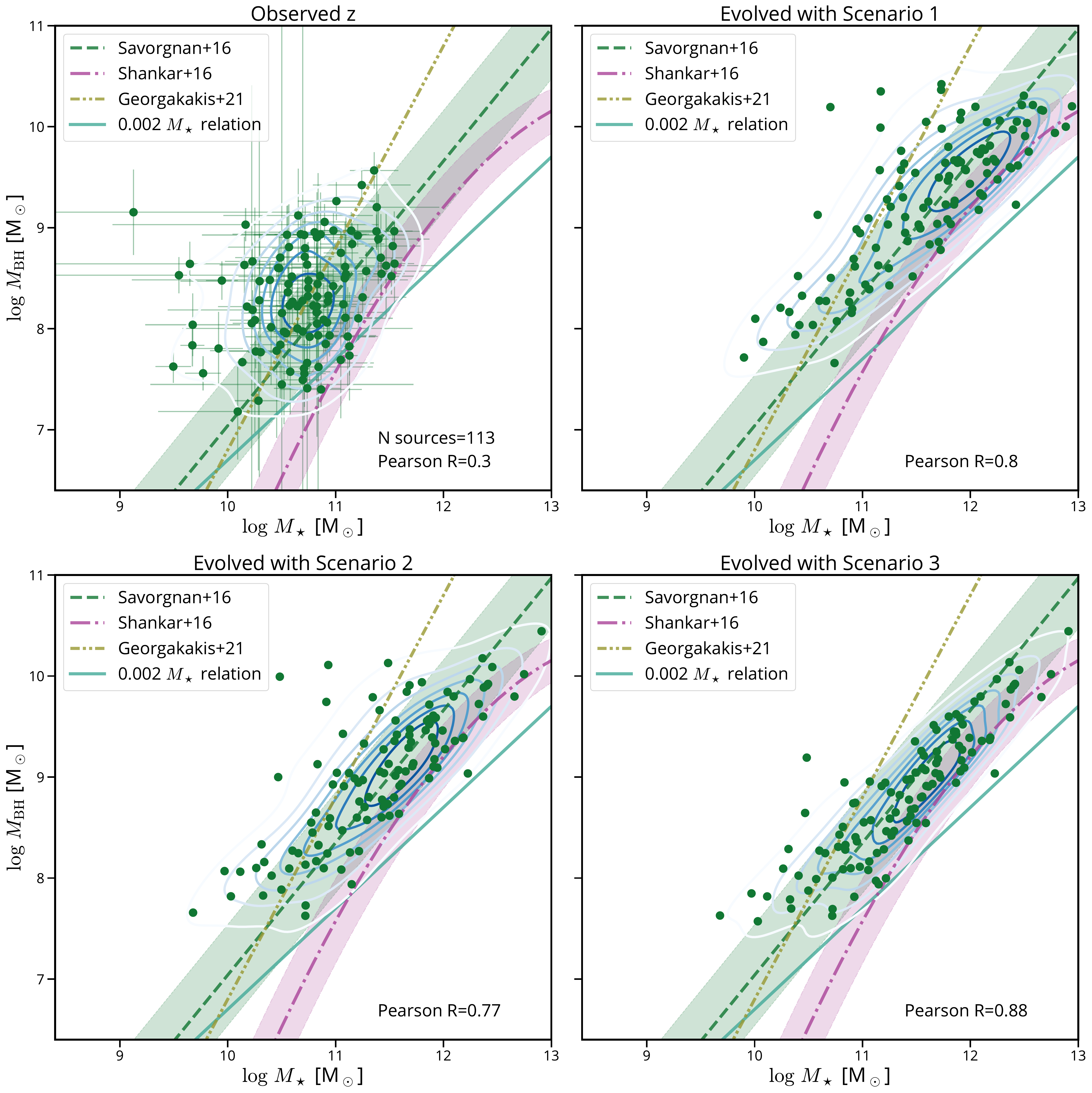}
   \caption{\textit{Upper left panel:} The observed $M_\mathrm{BH}$--$M_\star$ relation for all the \textbf{113} sources in our sample, at the observed redshift. Green solid dots are the value of masses for each source, with associated uncertainties, while contours are the distribution of these dots. In all panels, we show the expected local $M_\mathrm{BH}$--$M_\star$ scaling relations: in turquoise the $0.002~M_\star$ from \citet{2003ApJ...589L..21M}; in green the relation from \citet{2016ApJ...817...21S} with uncertanties at 1-$\sigma$; in purple the unbiased version from \citet{2016MNRAS.460.3119S} with uncertanties; and in brown the parametrization by \citet{2021MNRAS.508.3463G}. The other three panels show the forward modeling of our sources to \textit{z}\,$=$\,0 using different methods. \textit{Upper right panel:} the most basic model, with constant rates (Scenario 1). \textit{Lower left panel} the model with a variable rate following the SFH (Scenario 2). \textit{Lower right panel} the model with a variable rate following the SFH and the energy limit for the black hole accretion (Scenario 3). See Section 6 for details. 
   We also show the number of sources and the Pearson correlation value on each bottom right corner.}
   \label{fig:MM-Evol}
\end{figure*}

In the previous sections, we derived reliable stellar and BH masses for 113 miniJPAS sources with \textit{z}\,$<$\,2.5. These two quantities are compared, with the associated uncertainties, in the upper left panel of Figure~\ref{fig:MM-Evol}. For reference, we also included some local $M_\mathrm{BH}$--$M_\star$ scaling relations, like $M_\mathrm{BH}$\,$=$\,0.002~$M_\star$ used to define $\lambda_\mathrm{sBHAR}$ and based on \citet{2003ApJ...589L..21M}. We plot also the relation derived by \citet{2020NatAs...4..282S}, based on the BH masses measured from the velocity dispersions from \citet{2016ApJ...817...21S}. Since this method of measuring masses can have biases due to, e.g., the spatial resolution of the instrument used, we also included an unbiased relation proposed by \citet{2016MNRAS.460.3119S}. We also included the parametrization obtained by \citet{2021MNRAS.508.3463G}, where they presented an empirical model for BH populations in large cosmological volumes. 

Our sample does not seem to follow any of the local relations. Overall, a correlation between $M_\star$ and $M_\mathrm{BH}$ is not statistically significant with a Pearson correlation coefficient of only $\sim$\,0.3. This work is not the first to find overmassive BH compared with the local $M_\mathrm{BH}$--$M_\star$ scaling relation. For example, \citet{2020ApJ...888...37D} studied a sample of AGN between 1\,$<$\,\textit{z}\,$<$\,2, and they found $M_\mathrm{BH}$ almost three times more massive than those predicted by typical local $M_\mathrm{BH}$--$M_\star$ relations. It is clear that both physical effects as well as selection effects play a role in causing such apparent discrepancies. 

In any case, thanks to our comprehensive multiwavelength analysis, we also have reliable estimates of the SFR and the BHAR for all the sources in our sample.  
\citet{2010ApJ...708..137M} obtained stellar and black hole masses, and their SFR and BHAR for a sample similar in size ($\sim$\,100) of X-ray-selected AGN at \textit{z}$\sim$\,1.2. They showed that, assuming a constant value for the rates, an evolution of the sources for 300~Myr brings to a reduction of the scatter of the relation, and a new position in the $M_\mathrm{BH}$--$M_\star$ plane closer to the local scaling relations.

A simple constant model for the rates can be a valid assumption for evolution within short times (like nearby sources). Still, a constant value of SFR across longer times is likely an oversimplification. In fact, a galaxy will increase its stellar mass during its lifetime following a SFR that is not constant and that depend on the gas reservoir and its cooling time. This reservoir change over time by the effect of, for example, supernova, AGN feedback and/or mergers \citep{2004ApJ...600..580G,2006ApJ...650...42L,2007MNRAS.375.1189M,2020MNRAS.496.3943F}. A similar situation occurs for the SMBH mass. There are hints pointing out that the BH accretion rate shows a similar trend as the one observed in the star formation up to the peak epoch of galaxy-AGN coevolution (\textit{z}\,$\sim$\,3, e.g., \citealt{2014ARA&A..52..415M,2015MNRAS.451.1892A}). Nevertheless, this evidence is valid only for active black holes, where the accretion can be measured. To take into account the AGN duty cycle \citep{2014ApJ...782....9H}, a more complex approach is necessary, based on continuity equation approximations \citep[e.g.,][]{1992MNRAS.259..725S,2013MNRAS.428..421S}, as well as direct measurements of average BH accretion rates across large samples of active and inactive galaxies \citep[e.g.,][]{2018MNRAS.475.1887Y,2020A&A...642A..65C}.

We applied these basic ideas and performed forward modeling of our source properties to estimate the host galaxy stellar and BH masses that each source will have at \textit{z}\,$=$\,0, assuming an isolated evolution (merger-free). While mergers can be the main triggering mechanism of AGN activity for luminous sources \citep{2012ApJ...758L..39T,2018PASJ...70S..37G,2020A&A...637A..94G}, also observational studies and simulations suggest that mergers are not the dominant source of BH growth globally \citep{2015ApJ...810...74A,2018MNRAS.481..341S,2020MNRAS.494.5713M}, and even galaxies without major mergers since \textit{z}$\sim$\,1 follow the $M_\mathrm{BH}$--$M_\star$ relation \citep{2018MNRAS.476.2801M}. In any case, while our isolated evolution does not include the possibility of mergers, the measured accretion rate can be the result of a previous interaction of the host galaxy with their surroundings.

We considered three different evolving scenarios to model the late evolution of BHs and their host galaxies down to \textit{z}\,$=$\,0: (1) constant growth, (2) variable growth, and (3) variable growth with an energy limit.
Each model starts from the values of $M_\star$ and $M_\mathrm{BH}$ measured at the observed z and from the estimated rates (SFR and BHAR). Then, using a time step of 100~Myr, the model predicts the increment for both masses for each bin in time. 

The first scenario (Scenario 1) is the simplest one, where the rates are kept constant across time and are equal to the measured SFR and BHAR at the redshift of the sources up to \textit{z}\,$=$\,0. This approach is similar to what was considered in \citet{2010ApJ...708..137M}. 

The second scenario (Scenario 2) incorporates an evolution for both rates. For the SFR, we follow the SFH derived from the SED fitting. Instead of using analytical law for the BH growth (like Bondi accretion) and modeling the BH feedback, we choose a more straightforward approach. 
\citet{2015MNRAS.451.1892A} shows that SFR density is related to the BHAR until \textit{z}\,$=$\,3 for active galaxies, obtaining the BHAR from the X-ray luminosity function for AGN. Newer studies, like \cite{2018MNRAS.475.1887Y} and \cite{2020A&A...642A..65C}, show that this average relation is also valid for samples in different stages of their duty cycle (detected and not detected in X-ray). Since all this evidence suggests that the SFR's shape is similar to BHAR at least until \textit{z}\,$=$\,3 \citep{2015MNRAS.451.1892A}, we used the  $\tau$-delayed model of SFH of each source to obtain their BHAR at each time step. To scale the BH accretion history, we used the observed value of BHAR at observed \textit{z}.

The third scenario (Scenario 3) is similar to the second one, but a simple energy budget limits the BHAR. The BH is temporarily switched off if the total energy released by the AGN is larger than the gravitational binding of the host galaxy. This approach emulates a duty cycle and gives each source the possibility of periods of non-nuclear activity while the SF continues. While the shape of the relation is similar in studies for galaxies with diverse duty cycles \citep{2018MNRAS.475.1887Y} and for only active galaxies \citep{2015MNRAS.451.1892A}, the peak of the average BHAR at \textit{z}\,$=$\,2 for AGN is three times bigger. Our sources are X-ray detected, so the accretion activity is high at the observed \textit{z}. Thus, using these BHARs to model the accretion across history (as in scenario 2) can produce overmassive BHs; considering a duty cycle is fundamental for the evolution of each source (not averaged). In other words, each source has an energy budget and an active BH for the zero time step (equal to the observed \textit{z}). For the following time steps, if the energy limit allows it, the BH will be active, accrete, and follow the BHAR-SFR relation; if not, the BH will not have activity while the host galaxy will continue to form stars. This scenario better represents the normalization limits for the empirical BHAR-SFR relation for an X-ray-selected sample.

For Scenario 3, we calculated the energy released by the AGN at each time, as $\dot{E}_\mathrm{BH}(t) = \eta \epsilon c^2 \dot{M}_\mathrm{BH}(t)$, where $\eta$ is the radiative efficiency and $\epsilon$ is the coupling efficiency, and represents the energy fraction that couples with the surrounding medium. We used $\epsilon$\,$=$\,0.1 adopted as maximum value in \citet{2017MNRAS.465.3291W}. This value is also representative of other feedback models \citep{2018NatAs...2..198H}. For the gravitational binding energy, $E_\mathrm{gb}(t) = \frac{GM_{total}^2(t)}{r(t)}$, where $r$ is a radius representative of the galaxy. We took it from the relation between $M_\star$-size from \citet{2012MNRAS.422.1014I}. This relation is independent of redshift and type of galaxy, and the radius is calculated for each bin in time. We used $r$\,$=$\,$r_{90}$ since it contained 90\% of the light, and we are using all the source's light to estimate the $M_\star$. Because this, $M_{total}(t)$ takes into account the most massive components inside the $r_{90}$: $M_\star(t)$ and $M_\mathrm{halo}(t)$ up to that radius. $M_\star(t)$ is calculated at each bin. For $M_\mathrm{halo}(t)|_{r}$, we used the $M_\star$--$M_\mathrm{halo}$ relation from \citet{2020A&A...634A.135G}. This $M_\mathrm{halo}$ is the total mass of the halo, so we used a Jaffe profile\footnote{Other profiles with the same dependence were also tested, like the isothermal sphere, without significant changes. More variables are needed for more complex profiles, like the Einasto profile.} and the total mass to recover the fundamental parameters of the halo mass distribution and integrate $M_\mathrm{halo}(t)|_{r}$. After all the energies are calculated, we checked for each bin in time if  $E_\mathrm{BH}(t)$\,$\le$\,$E_\mathrm{gb}(t)$. If the energy is below the limit, the BH can accrete following the SFH. If not, the accretion stops and equals zero for that bin in time. The BH accretion can start again if the stellar mass increases because of high SFR; therefore, the gravitational binding is higher. 

The results of the forward modeling in the three scenarios described above are shown in \cref{fig:MM-Evol} (upper right, lower left and lower right, respectively). We performed a linear fit of the $M_\mathrm{BH}$--$M_\star$ relation evolved at \textit{z}\,$=$\,0, and in all three cases, the Pearson correlation coefficient factor increased substantially compared with the first panel. 

With the simplest model (Scenario 1), the sources seem to follow a relation of 0.01 between masses, with a correlation factor of $\sim$\,0.8. While this can be promising, the constant rates bring an estimate of $M_\star$ and $M_\mathrm{BH}$ considerably higher than those observed in the local Universe. While in the local Universe the highest values are $M_\mathrm{BH} \sim 10^{10} \mathrm{M}_\odot$ \citep{2021ApJ...921...36B} and $M_\star \sim 10^{12} \mathrm{M}_\odot$ \citep{2013AJ....145..101K}, we predict masses 10 times higher. 

Both BH and stellar masses distributions on the model following the Scenario 2 have slightly lower values than in Scenario 1. While $M_\star$ has values comparable to those observed in the local values, $M_\mathrm{BH}$ keeps being too high. 

The model with an energy limit for the BH accretion (Scenario 3) has the highest correlation coefficient factor (0.88) and masses closer to the local ones. Furthermore, our sample is closer to the dynamically-measure relation \citep{2016ApJ...817...21S} and unbiased relation \citep{2020NatAs...4..282S} than the relation obtained from forwarding model \citep{2021MNRAS.508.3463G}, with almost all its scatter inside the 1-sigma uncertantain of the dynamically-measured relation. With this scenario, we show that imposing an energy limit on the BH decouples the link between rates for specific periods of their evolution being critical to reproduce the $M_\mathrm{BH}$--$M_\star$ relationships. This limit is effortless to compute and still capable of lowering the BH masses at redshift 0, signaling that our simple assumptions are reasonable. 

Neither of the three scenarios was very sensitive to small changes in how $\dot{M}_\mathrm{BH}$ was derived. We ran them for the typical derivation $\dot{M}_\mathrm{BH}\propto L_\mathrm{AGN}$ without significant changes in the correlation values. We also applied the scenarios to 100 sets of different rates as a sanity check. We started from the same initial measured masses, but we evolved these sources with SFR, SFH, and $L_\mathrm{AGN}$ randomly sampled from a normal distribution centered in the mean value of our sample and with the same scatter. We found that for random samples, the evolution shows average correlation values significantly lower than our sample ($\sim$\,0.5). Therefore, for our forward modeling is essential to set actual input rates for each source and will not produce the same results for random samples.

The forward models explained in this Section are naive and straightforward by design, since they are based on a short set of observational inputs. The principal limitation is not having information about the gas reservoir of the galaxy; thus, accretion and star formation do not have a limit depending on how much gas is available. In this sense, our models are just a higher limit of how much the galaxy and its SMBH can grow. Nevertheless, we compared our results with a more complex evolution model, the Magneticum simulations\footnote{\smaller \url{http://www.magneticum.org}}. Magneticum is a set of fully hydrodynamical cosmological simulations that can trace structures through cosmic time, with different resolutions and box volumes \citep[details on][]{2014MNRAS.442.2304H,2015ApJ...812...29T}. In particular, we crossmatched our sources at observed redshift with the Magneticum/Box2, finding a similar stretch on the $M_\mathrm{BH}$--$M_\star$ relation at lower redshift on the evolved sources at \textit{z}\,$=$\,0.

\section{Summary and conclusions}
\label{sec:7}
We studied the host galaxy and central BH properties of a sample of X-ray-selected AGN using narrow-band data from the miniJPAS survey together with available multiwavelength data from UV to mid-infrared. We obtained robust parameters from SED fitting for 308 sources. We also measured BH masses from single-epoch spectra for a subsample of 113 sources. For this subsample, we provided reliable estimates of BH accretion rates and Eddington ratios. We also studied three different possible evolutionary scenarios for the subsample with BH mass estimates. We summarize below the main results of our work:

\begin{enumerate}
    \item The distribution of the Eddington ratio for our sample overgrows to lower values, peaking at $\lambda_\mathrm{Edd}\sim0.1$, and decreasing towards $\lambda_\mathrm{Edd}$\,$\sim$\,0.0001 (see the upper panel of Fig.~\ref{fig:Acrettion_ratios}). Since our sample is biased toward high-luminosity AGN, the distribution at $\lambda_\mathrm{Edd}$\,$<$\,0.01 needs to be studied with complete samples down to a luminosity of 10$^{42}$ erg s$^{-1}$ in the future. 
    \item We found that the distribution of Eddington ratios is on average about 0.6 dex smaller than its commonly used proxy $\lambda_\mathrm{sBHAR}$ (see \cref{fig:Acrettion_ratios} for a comparison). This difference must be studied in detail and highlights the importance of using high-quality photometric and spectroscopic data to derive physical parameters of accreting black holes.
    \item We derive accretion rates in physical units that depend on the expected radiative efficiency (see \cref{eq:dotmass} and \cref{fig:Acrettion}), using the estimated accretion luminosities and BH masses, obtaining less scatter than the accretion rates derived with a linear relation with $L_\mathrm{AGN}$.
    \item We do not find a correlation between the measured BH mass ($M_\mathrm{BH}$) and the galaxy stellar mass ($M_\star$) for the sources in our sample (upper left panel in Fig. \ref{fig:MM-Evol}).
    \item The fact that in our sample we measure overmassive BHs for their stellar masses compared to the local relations can be either ascribed to biases on $L_\mathrm{AGN}$ or to evolutionary effects. To test this second hypothesis, we applied forward modeling for our sources to the present time, considering three different scenarios for the growth history of both BH and host stellar masses. For all scenarios the $M_\mathrm{BH}$--$M_\star$ relation streches towards \textit{z}\,$=$\,0. 
    \item We found that the scenario that uses the SFH measured from the SED fitting, the SFR-BHAR relation, and an energy limit on the BH accretion as main hypotheses would evolve the sources to a $M_\mathrm{BH}$--$M_\star$ relation closer to the local one (see \cshref{sec:6} and lower right panel of \cref{fig:MM-Evol}).
    \item We cannot reproduce the diminution of scatter on the evolved $M_\mathrm{BH}$--$M_\star$ from observed $M_\mathrm{BH}$ and $M_\star$ and selecting random BHAR and SFR. Thus, the critical point is to start from the actual differential terms. Our evolution scenarios predict that sources below (above) the $M_\mathrm{BH}$--$M_\star$ relation will experience faster (slower) BH growth compared to galaxy build-up.
    \item The evolved $M_\mathrm{BH}$--$M_\star$ relation is consistent with the main relations observed for the local Universe. Our model evolves the galaxy in an isolated way following an empirical relation and without invoking the presence of mergers. All of these may witness a physical connection between integral and differential properties of the host galaxy and their central BHs and are essential to reproduce observations in the local Universe.
    \item The finding of `overmassive' and `undermassive' central BHs, compared with the $M_\mathrm{BH}$--$M_\star$, do not imply that $M_\mathrm{BH}$--$M_\star$ evolved with time. More evidence is needed to confirm the scatter being reduced across cosmic times to understand better the coevolution scenario. 
\end{enumerate}


This work also demonstrates the importance of having narrow-band, medium-deep photometry in the optical to characterize host galaxy properties of AGN at moderate to high redshift. With its full capability, JPAS will increase orders of magnitudes the size of samples to shred more light in the framework of galaxy evolution studies.

Finally we note that the biases of the present work depend mainly on the target selection bias for the cross-matched spectra, and the relatively low number of sources in our sample, as a result of only a tiny fraction of sky being covered by miniJPAS observations. In the future, combining extensive area X-ray surveys (like eROSITA) with the more extensive coverage of the sky by J-PAS will allow us to repeat the study with a much larger sample and implement other statistical techniques to understand better the importance of the host parameters in the accretion ratios. Recovering the black hole masses from the J-spectra is also possible for the brighter, more massive sources, and an excellent photo-z estimation for AGN will allow us to study the co-evolution scenario without needing a spectrum.

\begin{acknowledgements}
    This  action  has  received  funding  from  the  European  Union’s  Horizon 2020 research and innovation programme under Marie Skłodowska-Curie grant  agreement  No  860744 "Big Data Applications for Black Hole Evolution Sutdies"  (BID4BEST\footnote{\smaller \url{https://www.bid4best.org/}}). This paper has gone through internal review by the J-PAS collaboration. The color schemes used in this work are color-blind friendly from Paul Tol's Notes\footnote{\smaller \url{https://personal.sron.nl/~pault/}}. We also acknowledge the use of computational resources from the parallel computing cluster of the Open Physics Hub\footnote{\smaller \url{https://site.unibo.it/openphysicshub/en}} at the Physics and Astronomy Department of the University of Bologna. We also kindly thank Dr. Alison Coil and Dr. Christopher Willmer for sharing the MMT spectra. I.E.L.~thanks  E.~Marchesini for feedback on this work. A.L.~is partly supported by the PRIN MIUR 2017 prot. 20173ML3WW 002 ‘Opening the ALMA window on the cosmic evolution of gas, stars, and massive black holes’. 
    K.D.~acknowledges support by the COMPLEX project from the European Research Council (ERC) under the European Union’s Horizon 2020 research and innovation program grant agreement ERC-2019-AdG 882679.
    C.R.A.~acknowledges the projects ``Feeding and feedback in active galaxies'', with reference PID2019-106027GB-C42, funded by MICINN-AEI/10.13039/501100011033, ``Quantifying the impact of quasar feedback on galaxy evolution'', with reference EUR2020-112266, funded by MICINN-AEI/10.13039/501100011033 and the European Union NextGenerationEU/PRTR, and from the Consejer\' ia de Econom\' ia, Conocimiento y Empleo del Gobierno de Canarias and the European Regional Development Fund (ERDF) under grant ``Quasar feedback and molecular gas reservoirs'', with reference ProID2020010105, ACCISI/FEDER, UE.
    J.C.M.~acknowledges partial support from the Spanish Ministry of Science, Innovation and Universities (MCIU/AEI/FEDER, UE) through the grant PGC2018-097585-B-C22. J.C.M.~also acknowledges support from the European Union’s Horizon Europe research and innovation programme (COSMO-LYA, grant agreement 101044612). P.C.~acknowledges support from Conselho Nacional de Desenvolvimento Cient\'ifico e Tecnol\'ogico (CNPq) under grant 310555/2021-3 and from Funda\c{c}\~{a}o de Amparo \`{a} Pesquisa do Estado de S\~{a}o Paulo (FAPESP) process number 2021/08813-7.
    L.A.D.G. and R.M.G.D.~acknowledge financial support from the State Agency for Research of the Spanish MCIU through the `Center of Excellence Severo Ochoa' award to the Instituto de Astrof\'isica de Andaluc\'ia (SEV-2017-0709), and to the PID2019-109067-GB100. I.M.~acknowledges financial support from the State Agency for Research of the Spanish MCIU through the ‘Center of Excellence Severo Ochoa’ award to the Instituto de Astrofísica de Andalucía (SEV-2017-0709). I.M. is also supported by the Spanish Ministry of Economy and Competitiveness under grant no. PID2019-106027GB-C41. R.S.~acknowledges grant number 12073029 from the National Natural Science Foundation of China (NSFC). 
    R.A.D.~acknowledges support from the Conselho Nacional de Desenvolvimento Científico e Tecnológico -CNPq through BP grant 308105/2018-4, and the Financiadora de Estudos e Projetos - FINEP grants REF. 1217/13 - 01.13.0279.00 and REF 0859/10 - 01.10.0663.00 and also FAPERJ PRONEX grant E-26/110.566/2010 for hardware funding support for the J-PAS project through the National Observatory of Brazil and Centro Brasileiro de Pesquisas Físicas. L.S.J.~acknowledges the support from CNPq (308994/2021-3)  and FAPESP (2011/51680-6).
    \\
    Based on observations made with the JST/T250 telescope at the Observatorio Astrofísico de Javalambre (OAJ), in Teruel, owned, managed, and operated by the Centro de Estudios de Física del Cosmos de Aragón (CEFCA). We acknowledge the OAJ Data Processing and Archiving Unit (UPAD) for reducing and calibrating the OAJ data used in this work. Funding for the J-PAS Project has been provided by the Governments of Spain and Aragón through the Fondo de Inversión de Teruel, European FEDER funding and the Spanish Ministry of Science, Innovation and Universities, and by the Brazilian agencies FINEP, FAPESP, FAPERJ and by the National Observatory of Brazil. Additional funding was also provided by the Tartu Observatory and by the J-PAS Chinese Astronomical Consortium. Funding for OAJ, UPAD, and CEFCA has been provided by the Governments of Spain and Aragón through the Fondo de Inversiones de Teruel; the Aragón Government through the Research Groups E96, E103, and E16\_17R; the Spanish Ministry of Science, Innovation and Universities (MCIU/AEI/FEDER, UE) with grant PGC2018-097585-B-C21; the Spanish Ministry of Economy and Competitiveness (MINECO/FEDER, UE) under AYA2015-66211-C2-1-P, AYA2015-66211-C2-2, AYA2012-30789, and ICTS-2009-14; and European FEDER funding (FCDD10-4E-867, FCDD13-4E-2685). 
    Funding for the Sloan Digital Sky Survey (SDSS) has been provided by the Alfred P. Sloan Foundation, the Participating Institutions, the National Aeronautics and Space Administration, the National Science Foundation, the US Department of Energy, the Japanese Monbukagakusho, and the Max Planck Society. The SDSS Web site is http://www.sdss.org/. The SDSS is managed by the Astrophysical Research Consortium (ARC) for the Participating Institutions. The Participating Institutions are The University of Chicago, Fermilab, the Institute for Advanced Study, the Japan Participation Group, The Johns Hopkins University, Los Alamos National Laboratory, the Max-Planck-Institute for Astronomy (MPIA), the Max-Planck-Institute for Astrophysics (MPA), New Mexico State University, University of Pittsburgh, Princeton University, the United States Naval Observatory, and the University of Washington. Observations reported here were obtained at the MMT Observatory a joint facility operated by the Univesity of Arizona and the Smithsonian Institution. Funding for the DEEP2 Galaxy Redshift Survey has been provided by NSF grants AST-95-09298, AST-0071048, AST-0507428, and AST-0507483 as well as NASA LTSA grant NNG04GC89G.
\end{acknowledgements}

\bibliographystyle{aa}
\bibliography{biblio.bib}

\begin{appendix} 
\section{Comparison between mag.AUTO and mag.PSFCOR}
\label{appendix:phot}

\begin{figure}
    \resizebox{\hsize}{!}{\includegraphics{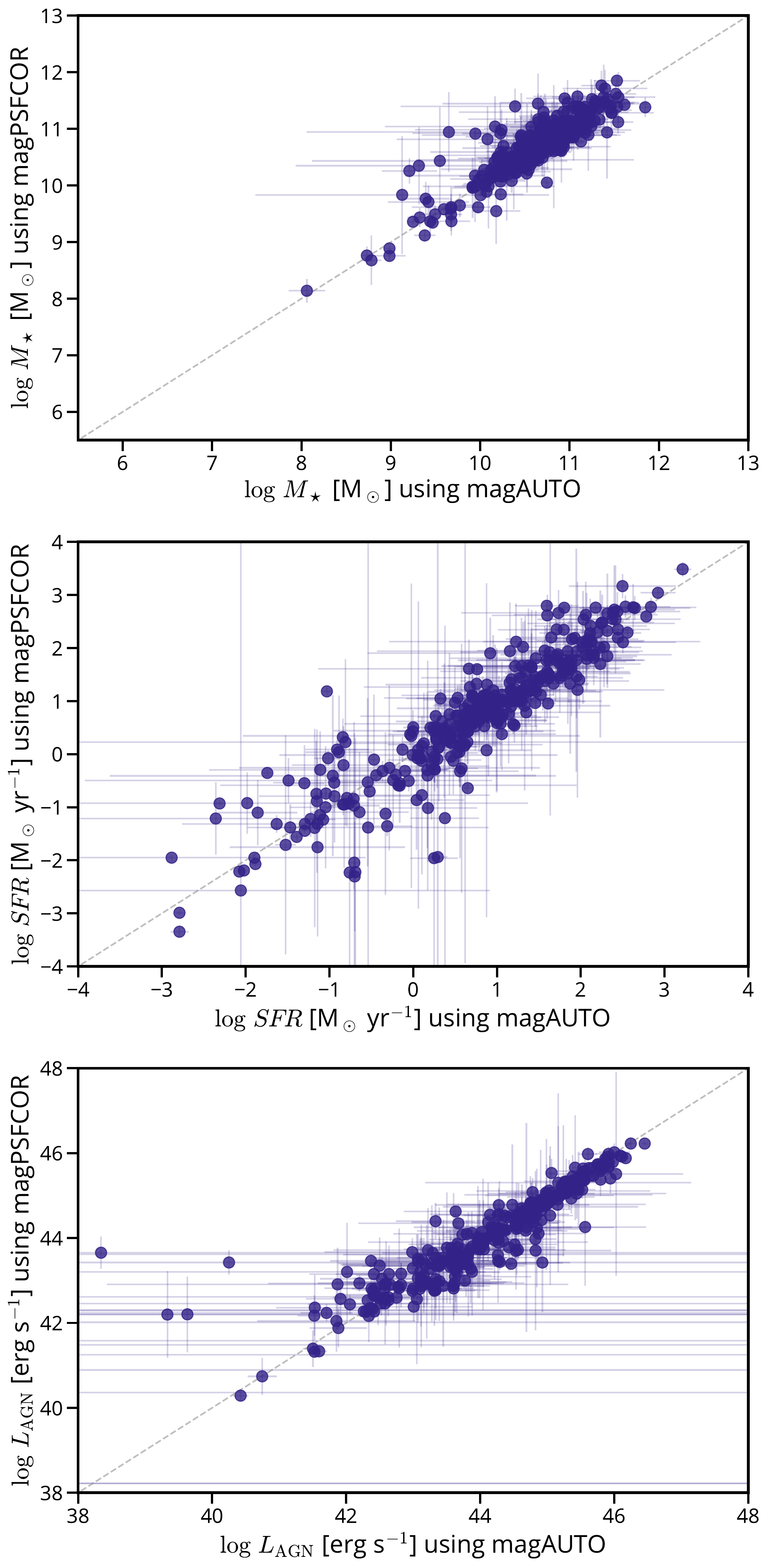}}
    \caption{Physical properties estimated obtained from the SED fitting, using different miniJPAS magnitudes as input. In all panel the 1:1 relation is the dashed line.}
    \label{fig:SED-comparing}
\end{figure}

In Fig \ref{fig:SED-comparing}, we show the comparison between physical parameters obtained from the SED fitting for both photometries. We used these parameters because they give basic information about the host galaxy ($M_\star$, SFR) and the AGN ($L_\mathrm{AGN}$). In each panel, the physical properties are in agreement for both photometries with the small mention that the relative errors of the \verb mag.PSFCOR \ is slightly higher than for \verb mag.AUTO. The median of the relative errors for \verb mag.AUTO  are $\tilde{\mathrm{RE}}_{\mathrm{M}_\star}$\,$=$\,0.16, $\tilde{\mathrm{RE}}_\mathrm{SFR}$\,$=$\,0.24 and $\tilde{\mathrm{RE}}_{\mathrm{L}_\mathrm{AGN}}$\,$=$\,0.26 while than for \verb mag.PSFCOR \ are 0.16, 0.33 and 0.27 respectively. The reported values are the bayesian parameters obtained from CIGALE after the exclusion explained in \cref{subsec:33}.

\end{appendix}
\end{document}